\documentclass{iopart}

\usepackage{graphicx}
\usepackage{bm}
\usepackage{sidecap}

\def\be{\begin{equation}}
\def\ee{\end{equation}}

\def\bea{\begin{eqnarray}}
\def\eea{\end{eqnarray}}

\def\l{\lambda}
\def\d{\delta}
\def\o{\omega}

\begin{document}

\title{Dynamics of the attractive 1D Bose gas:  analytical treatment from integrability}

\author{Pasquale Calabrese${}^{1}$ and Jean-S\'ebastien Caux${}^{2}$}

\address{$^{1}$Dipartimento di Fisica dell'Universit\`a di Pisa and INFN, 
Pisa, Italy}
\address{$^2$Institute for Theoretical Physics, Universiteit van Amsterdam,
1018 XE Amsterdam, The Netherlands}

\begin{abstract}
The physics of the attractive one-dimensional Bose gas (Lieb-Liniger model) is
investigated with techniques based on the integrability of the system.
Combining a knowledge of particle quasi-momenta to exponential precision in the system size 
with determinant representations of matrix elements of local operators 
coming from the Algebraic Bethe Ansatz, we obtain rather general analytical
results for the zero-temperature dynamical correlation functions of the
density and field operators.  Our results thus provide quantitative predictions for possible future
experiments in atomic gases or optical waveguides.
\end{abstract}

\maketitle

\contentsline{section}
 {\numberline {1}Introduction}{2}
\contentsline{section}
 {\numberline {2}Setup}{4}
\contentsline{section}
 {\numberline {3}Eigenstates}{5}
\contentsline{subsection}
 {\numberline {3.1}The ground state}{7}
\contentsline{subsection}
 {\numberline {3.2}Excited states}{7}
\contentsline{subsubsection}
 {\numberline {3.2.1}Single-particle states}{8}
\contentsline{subsubsection}
 {\numberline {3.2.2}Two-particle states}{8}
\contentsline{subsection}
 {\numberline {3.3}Norm of a general state, general remarks}{9}
\contentsline{section}
 {\numberline {4}Form factors}{10}
\contentsline{section}
 {\numberline {5}Density correlation function}{10} 
\contentsline{subsection}
 {\numberline {5.1}Sum rules}{11}
\contentsline{subsection}
 {\numberline {5.2}$N$ contribution (one-particle)}{11}
\contentsline{subsection}
 {\numberline {5.3}$N-M:M$ contribution (two-particle)}{12}
\contentsline{subsection}
 {\numberline {5.4}Static structure factor}{13}
\contentsline{section}
 {\numberline {6}One-body function}{14}
\contentsline{subsection}
 {\numberline {6.1}$N-1$ contribution (one-particle)}{15}
\contentsline{subsection}
 {\numberline {6.2}$N-1-M:M$ contribution (two-particle)}{15}
\contentsline{subsection}
 {\numberline {6.3}The momentum distribution function}{17}
\contentsline{section}
 {\numberline {7}Conclusions and perspective}{18}
\contentsline{section}
 {\numberline {Appendix A}\hspace{2cm}Reduced Bethe equations}{19}
\contentsline{section}
 {\numberline {Appendix B}\hspace{2cm}Norm of a general state}{20}
\contentsline{section}
 {\numberline {Appendix C}\hspace{2cm}Density operator form factor}{21}
\contentsline{section}
 {\numberline {Appendix D}\hspace{2cm}Field operator form factor}{23}
\contentsline{section}
 {\numberline {Appendix E}\hspace{2cm}Proof of summation formula}{24}

\section{Introduction}
The experimental realization of trapped quasi-one-dimensional atomic gases in optical lattices
during the last few years \cite{GoerlitzPRL87,GreinerPRL87,MoritzPRL91,ParededNATURE429,KinoshitaSCIENCE305,K2,K3,Bloch0704.3011}
has provided a new impetus for the study of the
effects of strong correlations on the physical properties of fundamental quantum-mechanical
systems of interacting particles.  One very interesting aspect of this still emerging field,
in contrast to traditional condensed matter physics, is that more-or-less ideal realizations
of famous toy models like the interacting Bose gas in one dimension (the Lieb-Liniger model
\cite{LiebPR130}) can be constructed and investigated.  It is moreover usually possible to
tune the parameters to arbitrary values, in contrast to the fixed values in a solid-state
crystal, and to obtain systems in which disorder and impurities are essentially absent.

Experimental measurements can typically be related to relatively simple theoretical objects 
like expectation values and correlation functions of local fields.  Ideally, one would like
to obtain direct quantitative comparisons between these, providing an extremely nontrivial
check on our understanding of the physics involved.  
From the theory side, calculating the response functions of a strongly correlated system
is more often than not a difficult, if not insurmountable task.  In one dimensional quantum
mechanics, the situation is not as hopeless as in higher dimensions in view of the
existence of many nonperturbative methods \cite{GiamarchiBOOK}.  While features such
as asymptotic correlation functions can be obtained from bosonization \cite{HaldanePRL47,GogolinBOOK},
it remains however extremely rare to be able to provide general analytical results valid for all
ranges of momentum and frequency.  

Our objective in this paper is to provide an extensive treatment of a situation
where much analytical progress is possible, namely that of the Lieb-Liniger model
for a one-dimensional Bose gas in the attractive regime.  
Like the repulsive case, the attractive one-dimensional gas is solvable with a 
Bethe Ansatz, so we can hope for full nonperturbative solutions.  
On the other hand, although the Bethe Ansatz offers a good basis for providing an 'exact' solution to
some models of interest in one-dimensional quantum physics, it is not necessarily
an easy framework to use to compute dynamical correlation functions or other objects
going beyond simple equilibrium thermodynamics.  In the case of the Lieb-Liniger model, 
while it is now possible to obtain
accurate numerical results for the dynamical structure factor \cite{CauxPRA74} and 
for the one-body function \cite{CauxJSTATP01008} for the {\it repulsive} Lieb-Liniger
model by mixing the Algebraic Bethe Ansatz with intensive numerical work, obtaining 
a full analytical solution for these objects in the thermodynamic limit 
remains a severe challenge \cite{KorepinBOOK}.  Here, we demonstrate that the {\it attractive} gas is
in fact fully tractable analytically. 
A summary of our results has appeared recently in \cite{CalabresePRL98}. 

The physics of the attractive Lieb-Liniger model, less conventional than in the
repulsive regime, has received attention in the past.
Most importantly, the ground state energy and wavefunction were investigated in 
\cite{McGuireJMP5,CalogeroPRA11}, and a more general treatment including excitations
was offered in \cite{CastinCRASP2}.  The purely one-dimensional case also serves as a basis
for the investigation of effects beyond the Gross-Pitaevskii mean-field in higher dimension
\cite{SalasnichJPB39}.  
Overall, the scattering and diffusion of solitons has also been studied \cite{SinhaPRL96,LeeEPL73},
and mean-field approaches based on the Gross-Pitaevskii equation have been developed
\cite{KavoulakisPRA67,KanamotoPRA67}.  
The case of attractive interactions is in principle
accessible experimentally in the context of bright solitons in quasi 1D
traps \cite{StreckerNATURE417,KhaykovichSCIENCE296,EiermannPRL92} 
since the effective one-dimensional coupling constant
\cite{InouyeNATURE392,OshaniiPRL81,BergemanPRL91} can effectively be tuned via
Feshbach resonance or transverse confinement to essentially any positive or negative value.
Such sort of solitonic physics can also be realized using photons in optical fibres
\cite{DrummondNATURE365}.

The paper is organized as follows.  We start by a reminder of the model, its solution in terms
of a Bethe Ansatz, and of the important correlation functions to consider in Section \ref{sec:setup}.
After describing eigenstates in more detail in Section \ref{sec:eigenstates}, we give the important
form factors in Section \ref{sec:FF}, which are then used in Section \ref{sec:DSF} and \ref{sec:1PGF} to
obtain the dynamical structure factor and one-body function at zero temperature.  Most
of the more tedious calculations have been relegated to \ref{app:Reduced_BE}-\ref{app:summation}.
We end by offering some conclusions and perspectives in Section \ref{sec:conclusion}.

\section{Setup}
\label{sec:setup}

The Hamiltonian of the Lieb-Liniger model is given by
\begin{eqnarray}
H = -\frac{\hbar^2}{2m} \sum_{j=1}^N \frac{\partial^2}{\partial {x_j^2}} 
+ 2 c \sum_{\langle i,j \rangle} \delta(x_i - x_j),
\label{LL}
\end{eqnarray}
where $\langle i,j \rangle$ represents the sum over all pairs, $c$ is 
the interaction parameter and $m$ the mass of the particles (atoms).  
In terms of experimental parameters the 1D coupling constant is 
$c=-\hbar^2/m a_{1D}$, where $a_{1D}$ is the effective 1D 
scattering length that can be tuned via Feshbach resonance or
transverse confinement \cite{InouyeNATURE392,OshaniiPRL81,BergemanPRL91}.
For definiteness, we consider a system of length $L$ (i.e. $0<x_i<L$) 
with periodic boundary conditions. 
From now on we fix $\hbar=2m=1$. 
In second quantization, this is nothing but the
nonlinear Schr\"odinger theory for a canonical Bose field, 
\begin{equation}
H = \int_0^L dx \left\{ \partial_x \Psi^{\dagger}(x) \partial_x \Psi(x) + c \Psi^{\dagger}(x) \Psi^{\dagger}(x) \Psi(x) \Psi(x) 
\right\}. \label{rhodef}
\end{equation}
We will be primarily interested in obtaining detailed results on
dynamical correlation functions of local operators of physical significance for the
attractive gas (\ref{LL}) with $c < 0$.  
We will first consider the density operator, which is defined
as
\begin{equation}
\rho (x) = \sum_{j=1}^N \delta(x-x_j) = \frac{1}{L} \sum_k e^{i kx} \rho_k=
\Psi^\dagger(x)\Psi(x). 
\end{equation}
More precisely, we will obtain its ground-state correlation function\footnote{Every time we write an operator with an explicit time dependence
  we obviously are working in the Heisenberg picture. Oppositely operators
  without time dependence (as in Eq. (\ref{rhodef})) are in the
  Schr\"odinger picture} 
\begin{equation}
S^\rho(x,t)=\langle\rho(x,t)\rho(0,0)\rangle, 
\label{DSF_xt}
\end{equation}
whose Fourier transform $S^\rho(k,\omega)$ is known as the dynamical structure 
factor (DSF).
In a similar way, we will also treat the case of the 
one-body dynamical correlation function of the field operator $\Psi(x,t)$,
\begin{equation}
S^\Psi(x,t)=\langle\Psi^\dag(x,t) \Psi(0,0)\rangle.
\label{DGF_xt}
\end{equation}

The dynamical structure factor (\ref{DSF_xt}) can at least in principle be obtained
experimentally by Fourier sampling of time-of-flight images \cite{DuanPRL96} or
Bragg spectroscopy \cite{StengerPRL82,StoeferlePRL92}.  The one-body function (\ref{DGF_xt}) is
obtainable through interference experiments
\cite{GritsevNATUREPHYSICS2,PolkovnikovPNAS103} or Raman spectroscopy
\cite{JaphaPRL82,LuxatPRL65,BlairBlackieNJP8,MazetsPRL94,DaoPRL98}, 
and its static version is the momentum distribution function which is 
directly accessed
using ballistic expansion \cite{KetterleREVIEW,ParededNATURE429}.

By introducing a resolution of unity, 
these correlation functions can be written in a Lehmann representation involving
a sum over intermediate states.  As a basis for the Fock space, we use the
eigenfunctions of the Hamiltonian as given by the Bethe Ansatz, which are combinations of plane waves
characterized by spectral parameters (rapidities) $\{ \lambda \}$:
\begin{eqnarray}
\Psi_N( x_1, ..., x_N | \lambda_1, ..., \lambda_N) = \prod_{N \geq j > k \geq 1} sgn(x_j - x_k) \times 
\nonumber \\
\sum_{P_N} (-1)^{[P]} 
e^{i \sum_{j=1}^N \lambda_{P_j} x_j + \frac{i}{2} \sum_{N \geq j > k \geq 1} sgn(x_j - x_k) \phi (\lambda_{P_j} - \lambda_{P_k})},
\label{BA_wavefn}
\end{eqnarray}
where $\phi (\lambda) = 2 \arctan ({\lambda}/{c})$.
These wavefunctions are symmetric in the position coordinates.  The rapidities of a 
given eigenstate obey a Pauli-like exclusion principle, namely the Bethe wavefunction 
vanishes if two rapidities coincide.  Quantization is achieved by enforcing periodic boundary
conditions.  It will become clear later on that our results are in fact insensitive to the
boundary conditions used.

Upon Fourier transformation ${\cal O} (k,\omega) = \int_0^L dx \int_{-\infty}^{\infty} dt e^{i(\omega t - kx)} {\cal O}(x,t)$,
the correlators we are interested in can be written as a sum over intermediate states labeled by $\mu$, 
\begin{eqnarray}
S^{\cal O}(k,\omega)= 2\pi L \sum_\mu \frac{|\Sigma^{\cal O}_{\mu \lambda^0}|^2}{||\lambda^0||^2 ||\mu||^2}
~\delta(\omega -E_\mu+E_{\lambda^0}),
\label{Lehmann}
\end{eqnarray}
where $\lambda^0$ denote the ground state rapidities, and the form factor (FF) $\Sigma^{\cal O}_{\mu \lambda} =\langle \mu|{\cal O}(0,0)|\lambda \rangle
=  \langle \mu | {\cal O}_{K_{\mu}- K_{\lambda}} | \lambda \rangle/L$ 
depends on the operator ${\cal O}$ and on the state $\mu$
(${\cal O}_K$ is the Fourier transform of the operator ${\cal O}$ at momentum $K$).
Throughout this paper, $|| \mu ||$ denotes the norm of state $\mu$. 

\section{Eigenstates}
\label{sec:eigenstates}
Let us begin by discussing the eigenstates, and by first recalling some standard
features of the Bethe Ansatz (\ref{BA_wavefn}) for the model at hand, and how they
adapt to the attractive case. 

As can be seen from the Schr\"odinger equation, 
given a set of rapidities, the energy and momentum of the corresponding state are
\begin{equation}
E_N = \sum_{j=1}^N \lambda_j^2, \hspace{1cm} P_N = \sum_{j=1}^N \lambda_j.
\end{equation}
Quantization of the theory is achieved by imposing the periodicity conditions
\begin{eqnarray}
\Psi (x_1, x_2, ..., x_N) = \Psi (x_2, ..., x_N, x_1 + L)
\end{eqnarray}
which lead to the Bethe equations 
\begin{equation}
e^{i\lambda_j L} = (-1)^{N-1} e^{-i \sum_k \phi (\lambda_j - \lambda_k)} = \prod_{k \neq j} \frac{\lambda_j - \lambda_k + ic}{\lambda_j - \lambda_k - ic}, 
\end{equation}
or in logarithmic form
\begin{equation}
\lambda_j L + \sum_k \phi (\lambda_j - \lambda_k) = 2\pi I_j, \hspace{0.5cm} j = 1, ..., N.
\label{BE_rep}
\end{equation}
The quantum numbers 
$I_j$ are half-odd integers if $N$ is even, and integers if $N$ is odd.

In the repulsive case, given a proper set of quantum numbers $\{ I \}$, 
the solution of the Bethe equations for the set of rapidities $\{ \lambda \}$ 
exists and is unique \cite{YangJMP10} due to the convexity of the 
Yang-Yang action associated with (\ref{BE_rep}).  
Furthermore all these solutions have real rapidities $\lambda_i$.
For the attractive case, the situation is completely different.  
We will define $\bar{c}=-c > 0$ as the interaction parameter;  let us rewrite our Bethe equations as
\begin{eqnarray}
e^{i \lambda_{\alpha} L} = \prod_{\beta \neq \alpha} \frac{\lambda_{\alpha} - \lambda_{\beta} - i\bar{c}}
{\lambda_{\alpha} - \lambda_{\beta} + i\bar{c}}.
\label{BE}
\end{eqnarray}
Consider now a complex rapidity $\lambda_{\alpha} = \lambda + i \eta$.  The Bethe equation
for this rapidity is
\begin{eqnarray}
e^{i\lambda_{\alpha} L} = e^{i \lambda L - \eta L} = \prod_{\beta \neq \alpha} \frac{\lambda_{\alpha} - \lambda_{\beta} - i\bar{c}}
{\lambda_{\alpha} - \lambda_{\beta} + i\bar{c}}.
\end{eqnarray}
We consider finite $N$ and $L \rightarrow \infty$.  If $\eta > 0$, we have $e^{-\eta L} \rightarrow 0$
on the left-hand side.  Looking at the finite product on the right-hand side, 
we conclude that there must thus be a rapidity $\lambda_{\alpha'}$ such that $\lambda_{\alpha'} = \lambda_{\alpha} - i \bar{c} + \mbox{O}(e^{-\eta L})$.
On the other hand, if $\eta < 0$, we have $e^{-\eta L} \rightarrow \infty$ on the left-hand side,
and there must thus be a rapidity $\lambda_{\alpha'}$ such that $\lambda_{\alpha'} = \lambda_{\alpha} + i \bar{c} + \mbox{O}(e^{-|\eta| L})$.

\begin{SCfigure}[40]
\includegraphics[width=5.5cm]{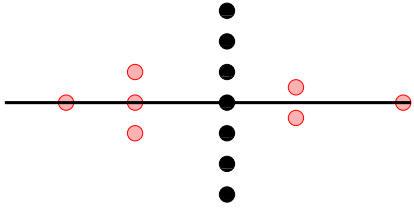}
\caption{Two string states of a gas of $N=7$ atoms. 
Black: The ground state consists of a single string centered at $k=0$ 
with all the $N$ particle aligned on the imaginary axis.
Red: An excited states with 4 strings of length $j=1,2,3$ 
and $N_1=2, N_2=1, N_3=1, N_{j>3}=0$.} 
\label{stringfig}
\end{SCfigure}

In general, the atoms will thus form bound states in the form of strings 
in the complex plane.
For a given number of atoms $N$, we can construct eigenstates with fixed string content by partitioning
$N$ into $N_j$ strings of length $j$, denoting the total number of strings as
$N_s$. We clearly have
\begin{eqnarray}
N = \sum_{j} j N_j, \hspace{1cm} N_s = \sum_{j} N_j.
\end{eqnarray}
Specifically, we will parametrize the string rapidities as
\begin{eqnarray}
\lambda_{\alpha}^{j, a} = \lambda_{\alpha}^j + i\frac{\bar{c}}{2} ( j + 1 - 2a) + i\delta_{\alpha}^{j,a},
\label{string_rap}
\end{eqnarray}
with exponentially small deviations $\delta \sim e^{-(cst)L}$
as usually done in these cases (see e.g. \cite{t-81,Tbook}).
In our string notation, the index $a = 1, \dots, j$ labels rapidities 
within the string, and $\alpha = 1, \dots, N_j$ labels strings 
of a given length.
A pictorial representation of two string states is given in 
Fig. \ref{stringfig}. 
We stress that perfect strings (i.e. with all the $\d_i=0$) are 
exact eigenstates in the limit $L\to\infty$ for arbitrary $N$.
It is then natural to consider the limit $L\to\infty$ at fixed $N$.
This is different from what done in the repulsive case where the limit 
$N,L\to\infty$ at fixed density $N/L$ is performed. Here, the $N$ particles
remain strongly correlated and bound to one another even when $L \to \infty$,
as we will explicitely see later on (see in particular the section about the 
norm of a string state \ref{normsec}).

The string deviations $\delta_{\alpha}^{j,a}$ are effectively studied 
in \cite{SakmannPRA72,Sykes0707.2422}, but
depend sensitively on the particular boundary conditions used.  We will however not need the explicit
values of these deviations:  we will only need the fact that they are small, and treat the resulting
limit carefully in the norm and form factor expressions.  For the moment, however, they can be dropped.

These bound states should be viewed as individual particles of mass $j$, with momentum and
energy of the string centered on $\lambda^j_{\alpha}$ given by
\begin{equation}
E_{(j, \alpha)} = j (\lambda^j_{\alpha})^2 - \frac{\bar{c}^2}{12} j(j^2 - 1), \hspace{1cm} 
P_{(j,\alpha)} = j \lambda^j_{\alpha}.
\end{equation}
Such strings are known but not commonly discussed in the literature on the Bose gas \cite{t-81,CastinCRASP2},
as they do not appear in the repulsive case.  However, their equivalents exist in integrable spin chains,
where they have been extensively studied.  The 'technology' to treat them, at least on the level of
eigenstates, is thus standard.  

The fact that all deviations are exponentially small means that we can obtain a much simpler set of
equations involving only the string centers $\lambda^j_{\alpha}$.  The derivation of these equations
in presented in \ref{app:Reduced_BE}, the final result being
\begin{eqnarray}
j \lambda_{\alpha}^j L - \sum_{(k, \beta)} \Phi_{jk} (\lambda_{\alpha}^j - \lambda_{\beta}^k) = 2\pi I_{\alpha}^j
\label{Reduced_BE}
\end{eqnarray}
with $I_{\alpha}^j$ half-odd integer (integer) if $N_j$ is even (odd), and where the scattering
phase shifts are
\begin{eqnarray}
\fl
\Phi_{jk} (\lambda) = (1 - \delta_{jk}) \phi_{|j-k|} (\lambda) + 2 \phi_{|j-k| + 2}(\lambda) + ... + 2\phi_{j+k - 2} (\lambda)
+ \phi_{j+k} (\lambda), 
\label{Reduced_Phi}
\end{eqnarray}
with $\phi_j (\lambda) = 2\arctan ({2\lambda}/{j \bar{c}})$.  
These strings are stable particles
under scattering with one another, and are therefore soliton-like objects.  One point worth emphasizing
is that the scattering phase shifts (\ref{Reduced_Phi}) are simply those of breathers in the classical
limit $\beta \rightarrow 0$ of the sine-Gordon model after a trivial reparametrization of the rapidity.  
The sine-Gordon soliton's mass is in this limit much higher than that of the breathers, which have
an evenly-spaced rest mass.  The string of order $j$
in the attractive Lieb-Liniger model is consequently not so much a 'soliton', but more accurately a
nonrelativistic sine-Gordon breather of order $j$.  

\subsection{The ground state}
The lowest energy state will be obtained by forming a bound state of all $N$ particles centered on zero
\cite{McGuireJMP5} (see Fig. \ref{stringfig}), namely by choosing
\begin{equation}
\lambda^{N,a} = i \frac{\bar{c}}{2} (N+1 - 2a) + \mbox{O}(\delta)\,.
\end{equation}
The corresponding energy is
\begin{eqnarray}
E_{GS} = \sum_a (\lambda^{N,a})^2 = -\frac{\bar{c}^2}{4} \sum_{a=1}^N (N+1 - 2a)^2 = -\frac{\bar{c}^2}{12} N(N^2 - 1).
\end{eqnarray}
Note that this goes like $N^3$, unlike the repulsive case where the energy is
not extensive.
Although our results are not limited to this case, we will often consider the limit of a 
large number of particles $N \gg 1$, with weak interactions $\bar{c}$ such that the parameter 
$g = \bar{c} N$ remains finite.  In this case, the ground state energy per particle also remains
finite, $E_{GS} = -{g^2}/{12}$.  We will find some similarly simplified limiting values for
the correlations functions.  It is however important to note that this is not a conventional
thermodynamic limit with finite energy density as in the repulsive case.

\subsection{Excited states}
Excitations above the ground state are then obtained by either simply giving
momentum to the ground state $N$ string, or more elaborately by partitioning it
into smaller strings to which individual momenta can be given.  We consider here
only the $N$ atom sector relevant for the dynamical structure factor.  The case
of states with $N-1$ (or less) atoms trivially follows.

We will label the string content of eigenstates by column-separated entries specifying
the length and number of each different string type.  For example
$N-M:M$ will be a state with a $N-M$ string and a $M$ string, and $N-M_1-2M_2:M_1:(M_2)_2$
a state with an $N - M_1 - 2M_2$ string, an $M_1$ string and two $M_2$ strings;  the
red state in Figure \ref{stringfig} is a $3:2:1_2$ state.

\subsubsection{Single-particle states}
Single-particle excited states will be obtained by giving finite momentum to the
ground state $N$-string, 
\begin{equation}
\mu^{N,a} = \mu + i \frac{\bar{c}}{2} (N+1 - 2a) + \mbox{O}(\delta).
\end{equation}
Such states have energy above the ground state given by
\begin{equation}
\omega_{N} (\mu) \equiv E_\mu-E_{GS}=N \mu^2 = k_{\mu}^2/N
\end{equation}
where $k_{\mu} = N \mu$ is the total state momentum.  
For these states, there is only one Bethe equation
for the string center $\mu$, namely $\mu = 2\pi {I}/{NL}$ with $I$ an 
integer, so that the momentum is quantized as for a free wave, 
$k_{\mu} = 2\pi {I}/{L}$.  In the limit
of large $N$, this energy band becomes flat and quasi-degenerate with the ground state.

\subsubsection{Two-particle states}
These are obtained by splitting up the ground-state $N$ string in two pieces.  In general,
consider having an $N-M$ and an $M$ string:
\begin{eqnarray}
\fl
\mu^{N-M,a} = \mu_s + i \frac{\bar{c}}{2} (N - M +1 - 2a) + \mbox{O}(\delta), \hspace{1cm} a = 1, ..., N-M, \nonumber \\
\fl
\mu^{M,a} = \mu_M + i \frac{\bar{c}}{2} (M + 1 - 2a) + \mbox{O}(\delta), \hspace{1cm} a = 1, ..., M.
\label{N-M:Mstrings}
\end{eqnarray}
The energy of this state above the ground state is given by
\begin{equation}
\omega_{N-M:M} (\mu_s, \mu_M) = \omega^0_{N-M:M} + (N-M) \mu_s^2 + M \mu_M^2, 
\label{N-M:Menergy}
\end{equation}
where we have defined the rest energy
\begin{equation}
\omega^0_{N-M:M} = \frac{\bar{c}^2}{4} N M (N-M).
\end{equation}
The total momentum is the sum of the two string momenta,
\begin{equation}
k = k_s + k_M = (N-M) \mu_s + M \mu_M,
\label{N-M:Mmomentum}
\end{equation}
so we can write the energy as 
\begin{equation}
\omega_{N-M:M} (k_s, k_M) = \omega^0_{N-M:M} + \frac{k_s^2}{N-M} + \frac{k_M^2}{M}.
\end{equation}
Similarly to the single-particle case, the Bethe equations are here very simple, namely
\begin{eqnarray}
(N-M) \mu_s L - \Phi_{N-M,M} (\mu_s - \mu_M) = 2\pi I_s, \nonumber \\
M \mu_M L + \Phi_{N-M,M} (\mu_s - \mu_M) = 2\pi I_M,
\end{eqnarray}
with $I_s, I_M$ integers.  In the limit of large L, we can thus ignore the scattering phase shift,
and take $\mu_s$ and $\mu_M$ as free parameters.  The total momentum $k$ of the state can take on
any value $2\pi I/L$, but the energy is bounded from below by 
\begin{equation}
\omega^l_{N-M:M} (k) = \omega^0_{N-M:M} + \frac{k^2}{N}.
\end{equation}
Given external frequency $\omega$ and momentum $k$ parameters, there are two solutions to the
dynamical constraints, namely
\begin{eqnarray}
\fl
\mu_s^{\pm}(k,\omega) &=& \frac{k}{N} \mp \left[\frac{M}{N(N-M)}\right]^{1/2} [\omega - \omega^l_{N-M:M}(k)]^{1/2}, \nonumber \\
\fl
\mu_M^{\pm}(k,\omega) &=& \frac{k}{N} \pm \left[\frac{N-M}{NM}\right]^{1/2} [\omega - \omega^l_{N-M:M}(k)]^{1/2}.
\label{N-M:MmusmuM}
\end{eqnarray}
Therefore, in the large $L$ limit, these states for a two-fold degenerate continuum 
beginning at the lower threshold $\omega^l_{N-M:M} (k)$ and extending to arbitrarily high energy,
\begin{equation}
\mbox{N-M:M continuum:} \hspace{1cm} \omega^l_{N-M:M}(k) \leq \omega < \infty.
\label{N-M:Mcontinuum}
\end{equation}
For finite $L$, this is of course not strictly a continuum:  only discrete energy levels $\omega$ then
exist, as determined from the Bethe equations.

\subsection{Norm of a general state, general remarks}
\label{normsec}
Before discussing form factors and correlation functions, we need to complete our
characterization of individual eigenstates by providing a formula for their norm.
Given a set of rapidities solution to the Bethe equations, the norm of 
the corresponding Bethe eigenstate is given by the Gaudin-Korepin formula \cite{GaudinBOOK,KorepinCMP86},
in this case
\begin{eqnarray}
||\Psi_N (\{ \lambda \}) ||^2 = |c|^N \prod_{j > k} \frac{\lambda_{jk}^2 + c^2}{\lambda_{jk}^2}
\mbox{Det}_N {\cal G}
\label{usual_norm}
\end{eqnarray}
where ${\cal G}$ is the Gaudin matrix whose entries are given by 
\begin{eqnarray}
{\cal G}_{jk} (\{ \lambda \}_N) = \delta_{jk} \left[ L + \sum_{l=1}^N K(\lambda_j, \lambda_l) \right] - K (\lambda_j, \lambda_k)
\label{Gaudin}
\end{eqnarray}
where the kernel is
\begin{equation}
K (\lambda, \mu) = \frac{2c}{(\lambda - \mu)^2 + c^2}.
\label{kernel}
\end{equation}

To cover the case of string states in the attractive regime, these formulas must be adapted
in view of its string deviation $\delta^{j,a}_{\alpha}/\delta^{j, a\pm1}_{\alpha}$ indeterminacy.  
We do this in detail in \ref{app:norm}, where we obtain the final result
\begin{eqnarray}
||\Psi_N (\{ \lambda_{\alpha}^j \}) ||^2 = (L \bar{c})^{N_s} \prod_{j} j^{2N_j} 
\prod_{(k, \beta) > (j, \alpha)} F_{jk}(\lambda_{\alpha}^j - \lambda_{\beta}^k)
\label{string_state_norm}
\end{eqnarray}
where
\begin{eqnarray}
F_{jk} (\lambda) = \frac{\lambda^2 + (\frac{\bar{c}}{2} (j + k))^2}
{\lambda^2 + (\frac{\bar{c}}{2} (j - k))^2}.
\end{eqnarray}

The norm of a single $N$ string is thus simply
\begin{equation}
|| \lambda ||^2 = \bar{c} L N^2
\end{equation}
and that of an $N-M:M$ state with rapidities $\mu_s, \mu_M$ respectively for the $N-M$ and $M$ strings,
\begin{equation}
\fl
|| \{ \mu_{s}, \mu_M \}||^2 = \bar{c}^2 L^2 (N-M)^2 M^2 \frac{(\frac{\mu_{s} - \mu_M}{\bar{c}})^2 + (\frac{N}{2})^2}
{(\frac{\mu_{s} - \mu_M}{\bar{c}})^2 + (\frac{N}{2} - M)^2}.
\end{equation}

As we already mentioned, unlike what happens in the repulsive case, where the low-density limit $N$ finite, $L \rightarrow \infty$
is trivial, the physics of the attractive case remains nontrivial.  This is not a low-density limit, as
the atoms tend to clump together in wavepackets of finite extent.  The ground state wavefunction is thus
really a localized wavepacket spread out uniformly over the whole system, as can be seen from our choice
of normalization:  this depends on $L$ generally as $L^{N_s}$, thus counting the number of strings
present, reflecting the fact that the strings are essentially independent, almost free particles.
In the presence of a weak harmonic confining trap in an experiment, the ground state would thus be
obtained by a convolution of the $N$ string wavepacket wavefunction with a ground state harmonic
oscillator part.  Our results could straightforwardly be adapted to cover this case.

\section{Form factors}
\label{sec:FF}
Since the Bethe wavefunctions are rather involved, comprising a factorially large number of
separate free wave terms, it is a priori extremely difficult to calculate matrix elements of
local operators in this basis.  However, the calculation can be performed for a number
of integrable models (including the Bose gas) using the technology of the Algebraic Bethe
Ansatz, which is sufficiently powerful to provide a handle on this combinatorial-like problem.
In fact, the ABA provides these matrix elements for the
density and field operators in terms of determinants of matrices whose entries are simple
analytical functions of the rapidities of both eigenstates involved \cite{SlavnovTMP79,SlavnovTMP82,KojimaCMP188,CauxJSTATP01008}.
These formulas are also applicable, at least in principle, to the attractive Bose gas,
provided some care is taken with string deviations.  The purpose of this section is to
perform this calculation.

For the repulsive gas, the form factors have to be evaluated numerically, since the
solutions to the Bethe equations cannot be found analytically.  For the attractive case,
however, we can actually calculate these determinants analytically to leading order in
the exponentially small string deviations, since we know the ground state to this precision
without having to solve any Bethe equation.

The form factors of the field and density operators between the ground state and
an arbitrary excited state are computed in \ref{app:density_FF} and \ref{app:field_FF}, and are given by
\begin{eqnarray}
|\Sigma_{\mu \lambda^0}^\rho|&=& \frac{P_{\mu}^2}{\bar{c}} N!(N-1)! \prod_{(j,\alpha)}
H_{j} ({\mu^j_{\alpha}}/{\bar{c}}), 
\label{FFs} \\
|\Sigma_{\mu \lambda^0}^\Psi|&=& \bar{c}^{\frac{1}{2}} N!(N-1)! \prod_{(j,\alpha)}
H_{j} (\mu^j_{\alpha}/\bar{c}), 
\label{FFg}
\end{eqnarray}
where we have defined the function
\begin{equation}
H_{M} (x) = \left| \frac{\Gamma (\frac{N - M}{2} + ix)}{\Gamma (\frac{N + M}{2} + ix)}\right|^2.
\end{equation}
For later convenience, the asymptotic properties of $H$ as $N \gg M$ are given by
\begin{eqnarray}
H_{M} (x) \longrightarrow \left[ \frac{N^2}{4} + x^2 \right]^{-M} (1 + {\cal O} (M/N)), \nonumber \\
H_{N-M} (x) \rightarrow \frac{F_M (x)}{F_{2N} (x)} \left[ N^2 + x^2 \right]^{M/2} (1 + {\cal O} (M/N)),
\end{eqnarray}
where we have defined the function
\begin{equation}
F_M (x) = \left|\Gamma \left(M/2 + ix \right) \right|^2.
\end{equation}
For $x \ll N$, we have further that 
$H_{N-M} (x) \rightarrow {N^M F_M(x)}/{\Gamma^2(N)} + \dots$.

\section{Density correlation function}
\label{sec:DSF}
Let us now apply the results previously obtained for states and form factors to the
calculation of the dynamical structure factor.  We first present the sum rules by which our
results can be checked, and then treat each important class of excited states separately.
We also consider how the equations obtained simplify in the limit of many weakly interacting
atoms.

\subsection{Sum rules}
The first sum rule is for the second density moment $\langle\rho^2\rangle$,
which is given by the Hellmann-Feynman theorem as 
\begin{equation}
\langle\rho^2\rangle=
-\frac{1}{L}\frac{\partial E_0}{\partial \bar{c}} 
= \frac{\bar{c} N (N^2 - 1)}{6L} = \frac{\bar{c}N^3}{6L} + \dots \,, 
\label{HF}
\end{equation}
and we must therefore have
\begin{equation}
\frac{1}{L} \sum_k \int_0^{\infty} \frac{d\omega}{2\pi} S^{\rho} (k, \omega)
= \langle \rho (0,0) \rho(0,0) \rangle = \frac{\bar{c} N (N^2 - 1)}{6L}.
\end{equation}
Another sum rule of practical interest is obtained from the first frequency moment at fixed momentum ($f$-sumrule):
\begin{equation}
f^{\rho} (k) = \int_0^{\infty} \frac{d\omega}{2\pi} \omega S^{\rho} (k, \omega) = \frac{N}{L} k^2.
\end{equation}
We will saturate these sum rules in what follows by adding up contributions to the structure factor 
coming from the lowest-lying bands of states.

\subsection{$N$ contribution (one-particle)}
When the rapidities $\{ \mu \}$ correspond to an $N$-string, 
we have from Eq. (\ref{FFs}) that
\begin{equation}
\fl
|\Sigma^{\rho}_{\mu \lambda^0}| = 
\frac{\Gamma (i\frac{\mu}{\bar{c}}) \Gamma (-i\frac{\mu}{\bar{c}}) \Gamma(N+1) \Gamma(N)}
{\Gamma (N + i\frac{\mu}{\bar{c}}) \Gamma (N -i\frac{\mu}{\bar{c}})} \frac{P_{\mu}^2}{\bar{c}} = 
\frac{N \Gamma^2(N) P_{\mu}^2/\bar{c}}{\prod_{a=1}^N \left[ (\mu/\bar{c})^2 
+ (a-1)^2 \right]}
\end{equation}
and $k \equiv P_{\mu} = N \mu$.  
The norm of this state is $|| \{ \mu \} ||^2 = \bar{c} L N^2$, so 
\begin{eqnarray}
S^{\rho}_N (k, \omega) = \frac{2\pi N^2}{L} \frac{\delta (\omega - k^2/N)}
{\prod_{a=1}^{N-1} \left[1 + (\frac{k}{\bar{c} N a})^2\right]^2}.
\end{eqnarray}
This contribution is thus a single peak centered on the $N$ string dispersion relation.

At large $N$ with $g = \bar{c} N$ fixed, this becomes 
\begin{equation}
S^{\rho}_N (k, \omega) = 2\pi \frac{N^2}{L} \frac{(\pi k/g)^2}{\sinh^2 \pi k/g} \delta (\omega - \frac{k^2}{N}).
\label{SNN}
\end{equation}

The density moment sum rule contribution from these intermediate states is for large $N$:
\begin{equation}
\frac{1}{L} \sum_k \int_0^{\infty} \frac{d\omega}{2\pi} S^{\rho}_N (k, \omega) \rightarrow
\frac{N^2}{L} \int_{-\infty}^{\infty} \frac{dk}{2\pi} \frac{(\pi k/g)^2}{\sinh^2 (\pi k/g)} 
= \frac{\bar{c} N^3}{6L}
\end{equation}
so these states completely saturate this sum rule for large $N$.

The $f$-sumrule contribution is simply written as
\begin{eqnarray}
f_N^{\rho} (k) = \int_0^{\infty} \frac{d\omega}{2\pi} \omega S^{\rho}_N (k, \omega) = \frac{N}{L} \frac{k^2}
{\prod_{a=1}^{N-1} \left[1 + (\frac{k}{\bar{c} N a})^2\right]^2}
\end{eqnarray}
which for large $N$ and fixed $g$ becomes
\begin{eqnarray}
f_N^{\rho} (k) = \frac{N}{L} k^2 \left[ \frac{\pi k/g}{\sinh \pi k/g}\right]^2 < \frac{N}{L} k^2.
\label{fSRN}
\end{eqnarray}
These states therefore do not completely saturate the $f$-sumrule, so we consider higher excited
states in order to find the missing part.

Note that if we instead considered $N\to\infty$ at fixed $\bar{c}$, the
$f$-sumrule would have been saturated by these states, i.e. the dynamics would
have been less interesting.

\subsection{$N-M:M$ contribution (two-particle)}
The next simplest excited states to consider are the two-particle $N-M:M$ states with $M = 1, 2, ...$.
The rapidities of these are parametrized using (\ref{N-M:Mstrings}), and the energy and momentum
with respect to the ground state are respectively given by (\ref{N-M:Menergy}) and (\ref{N-M:Mmomentum}).
Given an external frequency $\omega$ and momentum $k$ within the continuum (\ref{N-M:Mcontinuum}),
there exist two solutions to the dynamical constraints for the rapidities $\mu_s, \mu_M$ given
by equation (\ref{N-M:MmusmuM}).

From (\ref{FFs}), we find that the density operator form factors relevant to this class of
states are
\begin{equation}
|\Sigma^{\rho}_{\mu \lambda^0}| = N \Gamma^2(N) \frac{P_{\mu}^2}{\bar{c}}
H_{N-M}(\mu_{s}/\bar{c}) H_{M}(\mu_M/\bar{c}).
\end{equation}
Their contribution to the dynamical structure factor is thus
\begin{eqnarray}
S^{\rho}_{N-M:M} (k, \omega) = 2\pi L \sum_{\mu_s,\mu_M} \frac{|\Sigma^{\rho}_{\mu \lambda^0}|^2}{||\mu||^2 ||\lambda^0||^2} \delta_{k, K_{\mu}}
\delta (\omega - E_{\mu} + E_{\lambda^0}).
\end{eqnarray}
Let us explicitly rewrite the right-hand side as a function of the variables $k, \omega$.
The energy delta function can be rewritten using (\ref{N-M:MmusmuM}) as a function of $\mu_s$,
\begin{equation}
\fl
\delta (\omega - E_{\mu} + E_{\lambda_0}) = \left[\frac{M}{N(N-M)}\right]^{1/2}
\frac{\Theta(\omega -  \omega^l_{N-M:M}(k))}{2\left[\omega - \omega^l_{N-M:M}(k)\right]^{1/2}}
\sum_{\sigma = \pm} \delta (\mu_s - \mu_s^{\sigma}(k,\omega)).
\end{equation}
For large $L$, the summation over $\mu_s$ can be replaced by an integral, $\sum_{\mu_s} \rightarrow L\int_{-\infty}^{\infty} \frac{d\mu_s}{2\pi}$.
$\mu_M$ is then completely fixed, finally yielding
\begin{eqnarray}
\fl
S^{\rho}_{N-M:M} (k,\omega) = \frac{\Theta(\omega -  \omega^l_{N-M:M}(k))}{\left[\omega - \omega^l_{N-M:M}(k)\right]^{1/2}}
\frac{\Gamma^4(N) k^4}{2 N^{1/2}M^{3/2}(N-M)^{5/2} \bar{c}^5 L} \times \nonumber \\
\hspace{4cm}\times \sum_{\sigma = \pm}
\frac{H_{N-M}^2 (\mu_s^{\sigma}/\bar{c}) H_{M}^2 (\mu_M^{\sigma}/\bar{c})}{F_{N-M,M} (\mu_{s}^{\sigma} - \mu_M^{\sigma})}.
\end{eqnarray}
As everywhere, this expression is correct to leading order in the
exponentially small string deviations for any value of $N$ and $M$.
Each choice of $M$ gives a distinct continuum.

In the limit $N \gg M \sim 1$, this becomes
\begin{eqnarray}
\fl
S^{\rho}_{N-M:M} (k,\omega) = \frac{\Theta(\omega -  \omega^l_{N-M:M}(k))}{\left[\omega - \omega^l_{N-M:M}(k)\right]^{1/2}} \times \nonumber \\
\times 
\frac{k^4}{2 N^{2M+3} M^{3/2} \bar{c}^5 L} \sum_{\sigma = \pm} \frac{F_M^2(\mu_s^{\sigma}/\bar{c})}
{\left[ 1/4 + (\mu_M^{\sigma}/\bar{c}N)^2\right]^{2M}}.
\end{eqnarray}
The hierarchy is such that higher $M$ parts are suppressed by increasing
powers of $N$\footnote{If $M$ is of the order of $N$ it is straightforward to
  show that the contributions are exponentially suppressed for large $N$}.
The leading $M = 1$ part can be further simplified to
\begin{equation}
S^{\rho}_{N-1:1} (k,\omega) = \frac{N}{L} \frac{k^4}{2 g} \frac{\Theta(\omega - g^2/4)}
{[\omega - g^2/4]^{{1}/{2}} \omega^2} \sum_{\sigma=\pm} F_1^2 (x_{\sigma})
\label{S_rho_Nm1_1}
\end{equation}
with $F_1(x) = {\pi}/{\cosh(\pi x)}$ and $x_{\pm} = ([\omega -
g^2/4]^{\frac{1}{2}} \pm k)/g$.

\begin{figure}
\begin{tabular}{cc}
\includegraphics[width=6cm]{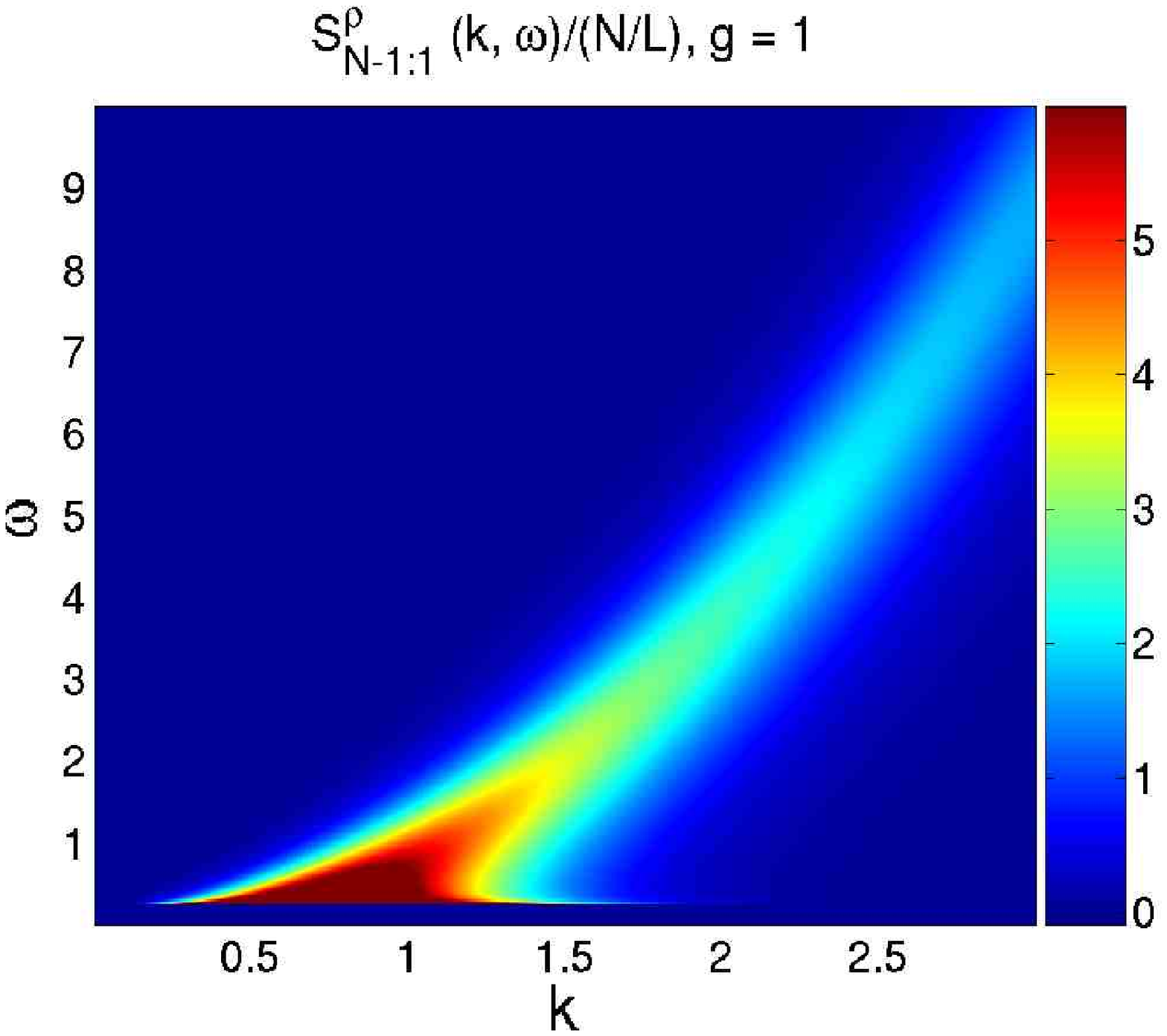} &
\includegraphics[width=6cm]{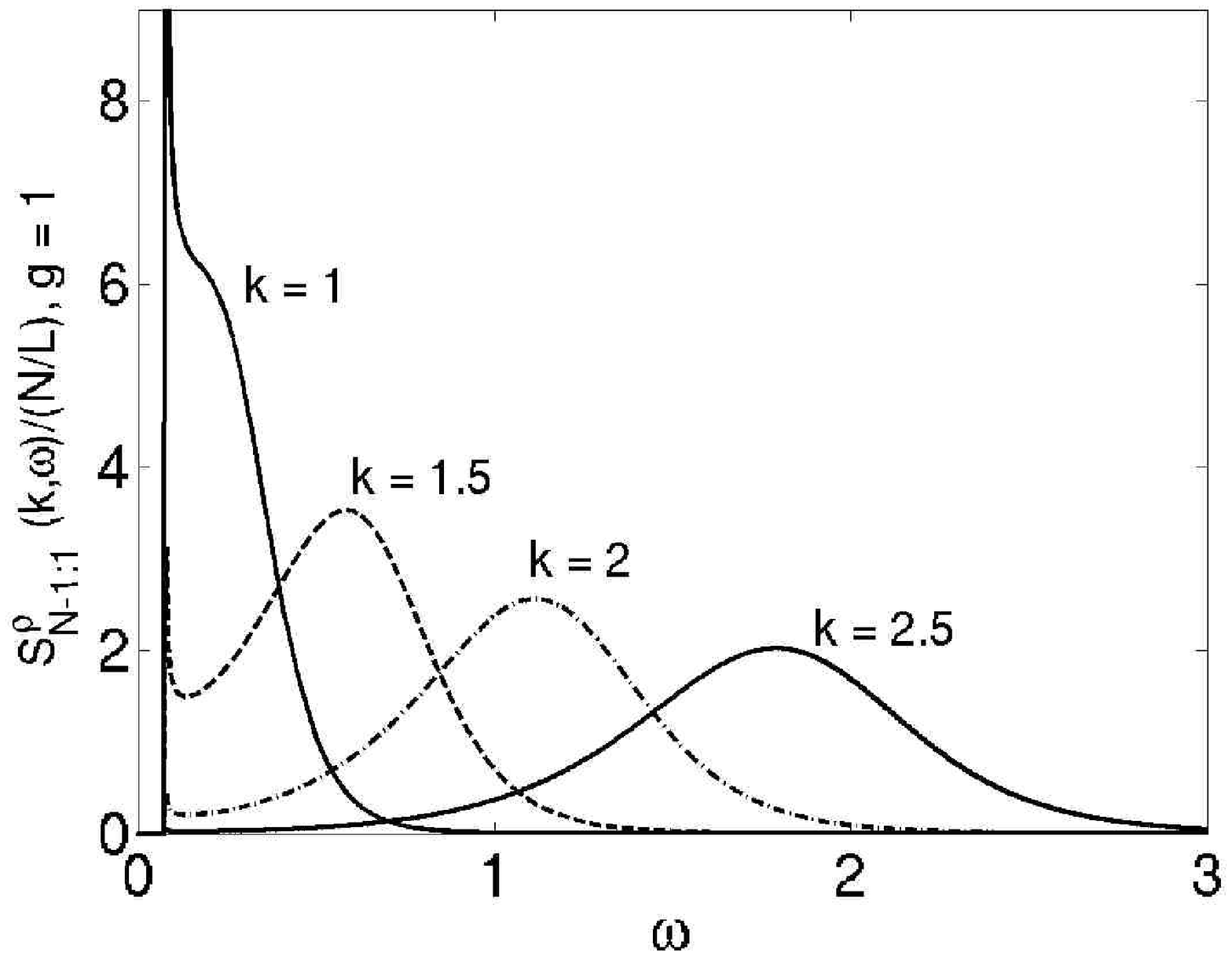} 
\end{tabular}
\caption{Left: contribution to the DSF coming from $N-1:1$ (two-particle) states,
in the large $N$ limit for $g = 1$.  The square-root singularity at the lower
threshold is accompanied by a maximum around $k = [\omega - g^2/4]^{\frac{1}{2}}$.
Right:  Fixed momentum cuts of Fig. \ref{Srhofig}, showing the weakening of
the square-root singularity at higher momentum, and the displacement of the
maximum towards higher energies.}
\label{Srhofig}
\end{figure}

The contribution of these state to the $f$ sum rule is written as the integral
(introducing $g \zeta = \sqrt{\omega - g^2/4}$)
\begin{equation}
f^{\rho}_{N-1:1}(k) =\frac{N}{L} \frac{\pi^2 k^4}{g^3} \int_{-\infty}^{\infty}
\frac{d\zeta}{2\pi} \frac{1}{\zeta^2 + 1/4} \frac{1}{\cosh^2 \pi(\zeta - k/g)}.
\end{equation}
In the upper half-plane, the first piece of the integrand has a single pole at $\zeta = i/2$ and 
the second double poles
at $\zeta = {k}/{g} + i (n + 1/2)$, $n = 0, 1, ...$.  Writing out the 
residues gives a series which
can be explicitly resummed, leading to the final result
\begin{equation}
f^{\rho}_{N-1:1}(k) =\frac{N k^2}L  \left(1 - \frac{(\pi k/g)^2}{\sinh^2(\pi k/g)} \right).
\end{equation}
These states therefore completely saturate the remaining part of the $f$-sumrule after
the $N$ string contributions (\ref{fSRN}) have been taken into account.  Higher frequency moments further
suppress the $S^{\rho}_N$ part;  in general, any frequency moment can be accurately obtained by approximating
the dynamical structure factor with
$S^{\rho}(k,\omega) = S^{\rho}_{N} (k,\omega) + S^{\rho}_{N-1:1} (k,\omega)$.

The two-particle ($N-1:1$) part of the DSF is plotted in the left panel of 
Figure \ref{Srhofig} for large $N$, with $g = 1$.  At the lower threshold, 
the DSF diverges as $(\omega - g^2/4)^{-{1}/{2}}$.
For $\o>g^2/4$  it is a monotonous decreasing function of $\o$
as long as $k/g<x_c=1.0565\dots$, 
whereas for $k/g>x_c$ it shows a broad peak, 
whose position grows like $k^2$ for large $k$ (its amplitude decreases 
like $k^{-1}$).
Away from this peak it decays exponentially.  
The right panel of Figure \ref{Srhofig} provides fixed momentum cuts showing these features in more detail.  

\subsection{Static structure factor}
As a byproduct of the previous results we can simply obtain the static
structure factor 
\begin{equation}
S^{\rho} (k) = \int_0^{\infty} \frac{d\omega}{2\pi} S^{\rho}(k, \omega)
\label{static_SF}
\end{equation}
from our results.  The contribution of one particle states is given by the  integral of 
Eq. (\ref{SNN}) over $\omega$, yielding
\begin{equation}
S^{\rho}_N (k) = \frac{N^2}{L} \frac{(\pi k/g)^2}{\sinh^2 (\pi k/g)}.
\label{SrhoN}
\end{equation}

\begin{figure}
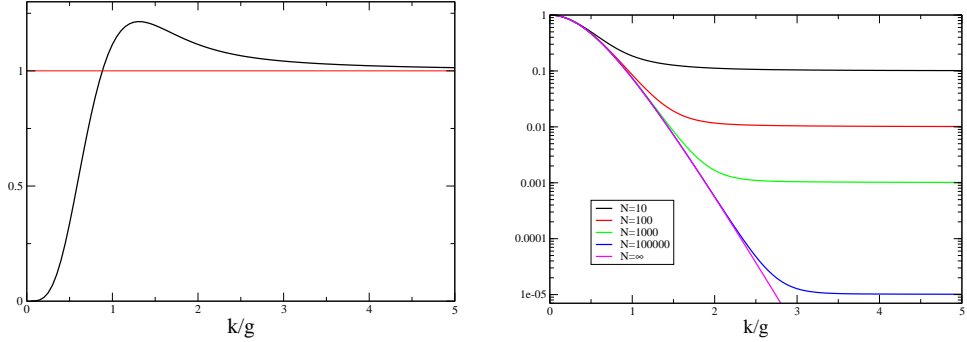

\begin{tabular}{cc}
\includegraphics[width=6cm]{staticDD.eps} &\hspace{2mm}
\includegraphics[width=6cm]{DDcross.eps}
\end{tabular}
\caption{The static structure factor.
Left: Two-particle contribution $S^{\rho}_{N-1:1} (k)/(N/L)$.  
Right:  One- plus two-particle contribution 
$(S^{\rho}_N(k)+S^{\rho}_{N-1:1} (k))/(N^2/L)$
for $N=10, 100, 1000,100000$ showing the crossover at $k/g\sim \ln N$.}
\label{staticfig}
\end{figure}

We can also evaluate the contribution of the $N-1:1$ states: 
\begin{equation}
S^{\rho}_{N-1:1} (k) = \frac{N}{L} \pi^2 (k/g)^4 \int_{-\infty}^{\infty}
\frac{d\zeta}{2\pi} \frac{1}{(\zeta^2 + 1/4)^2} \frac{1}{\cosh^2 \pi(\zeta - k/g)}.
\end{equation}
Once again, the residues series can be explicitly resummed, giving
\be
S^{\rho}_{N-1:1}(k)= \frac{2Nk^4}{Lg^4} \left[\frac{g^2}{k^2}
-\frac{\pi^2}{\sinh^2 \pi \frac{k}{g}}-{\rm Re} ~\psi_2\!\left(i \frac{k}{g}\right)\right]
\ee
where $\psi_2(z)$ is the polygamma function (second 
derivative of the logarithm of the Gamma function).
A plot of $S^{\rho}_{N-1:1}(k)$ is shown in the left panel of Fig.
\ref{staticfig}. It shows a broad peak at $k/g\sim1.2$. 
This is similar to what observed numerically in the super-Tonks-Girardeau 
gaslike regime in Ref. \cite{abcg-05}.
Although any eventual connection between the two results is far from being 
clear, it is tempting to believe that the quench ``experiment'' used to
realize the  super-Tonks-Girardeau gas has selected a two-particle state, 
in analogy to what claimed in \cite{bat-05}.

The one-particle contribution $S^{\rho}_N(k)$ for large $N$ grows like $N^2$,
and so it will always dominate against the two-particle one (growing like $N$)
for any $k$ at large enough $N$. 
However for large $k/g$, $S^{\rho}_N(k)$ decays
exponentially, whereas $S^{\rho}_{N-1:1}(k)$ goes to the constant value $N/L$.
Consequently, a crossover between the two regimes 
is expected to take place at relatively low momentum even for very large $N$. 
In fact, imposing $S^{\rho}_N(k)\sim N/L$ we have 
$k_c/g \sim (\ln N)/2\pi$. This crossover is explicitely shown in the right
panel of Fig. \ref{staticfig}.

\section{One-body function}
\label{sec:1PGF}
For the dynamical one-body function $S^{\Psi}(k, \omega)$, the relevant 
intermediate states are those made of $N-1$ atoms.  
When calculating the energy difference of a given excited state
with the ground state, we will thus have to take the chemical potential
$E^0_N - E^0_{N-1} = -{\bar{c}^2} N(N-1)/4$ into account.

The relevant sum rule here is very simple, namely that we should recover the particle density
by integrating over energy and summing over momenta,
\begin{equation}
\frac{1}{L} \sum_k \int_0^{\infty} \frac{d\omega}{2\pi} S^{\Psi}(k, \omega) =
\frac{N}{L}.
\label{opsumrule}
\end{equation}

\subsection{N-1 contribution (one-particle)}
In this case, we write the excited state as an $N-1$ string,
\begin{equation}
\mu^{N-1,a} = \mu + i\frac{\bar{c}}{2} (N - 2a), \hspace{1cm} a = 1, ..., N-1.
\end{equation}
The total momentum of this state is $P = (N-1) \mu$, and its energy above the ground state
(taking the chemical potential into account) is $(N-1) \mu^2$.  Its norm is 
$||\mu||^2 = \bar{c} L (N-1)^2$, and its field operator form factor with the ground
state is obtained from (\ref{FFg}) as
\begin{equation}
|\Sigma^{\Psi}_{\mu \lambda^0}| = \frac{\sqrt{\bar{c}} N \Gamma^2(N)}{\prod_{a=1}^{N-1}
\left[ (\mu/\bar{c})^2 + (a - 1/2)^2 \right]}.
\end{equation}
The one-body function contribution is thus
\begin{equation}
S^{\Psi}_{N-1} (k, \omega) = \frac{2\pi}{\bar{c}L} \frac{\delta(\omega - \frac{k^2}{N-1})}
{\prod_{a=1}^{N-1} \left[ (1 - \frac{1}{2a})^2 + (\frac{k}{\bar{c}Na})^2 \right]^2}.
\label{SPsi_Nm1}
\end{equation}
In the large $N$ limit at $g$ constant, this becomes
\begin{equation}
S^{\Psi}_{N-1} (k, \omega) = 
\frac{2\pi^3 N}{gL} \frac{\delta(\omega - k^2/N)}{\cosh^2 (\pi k/g)}.
\end{equation}
Integrating this over frequency and momenta, we find that
\begin{equation}
\frac{1}{L} \sum_k \int_0^{\infty} \frac{d\omega}{2\pi} S^{\Psi}_{N-1} (k, \omega) = \frac{N}{L}
\end{equation}
so these states completely saturate the integrated intensity sum rule in this limit.
Since we can also be potentially interested in higher frequency moments, which would then not
be completely saturated by this function, we consider higher excited states, the leading ones
being the two-particle ones.

\subsection{N-1-M:M contribution (two-particle)}
In complete parallel to the $N-M:M$ states considered for the structure factor, we here
consider the $N-1$ atom states with defined by having an $N-M - 1$ and an $M$ string:
\begin{eqnarray}
\fl
\mu^{N-M-1,a} = \mu_s + i \frac{\bar{c}}{2} (N - M - 2a) + \mbox{O}(\delta), \hspace{1cm} a = 1, ..., N-M-1, \nonumber \\
\fl
\mu^{M,a} = \mu_M + i \frac{\bar{c}}{2} (M + 1 - 2a) + \mbox{O}(\delta), \hspace{1cm} a = 1, ..., M.
\label{N-M-1:Mstrings}
\end{eqnarray}
The energy of this state above the ground state (again taking the chemical potential into account)
is given by
\begin{equation}
\omega_{N-M:M} (\mu_s, \mu_M) = \omega^0_{N-M-1:M} + (N-M-1) \mu_s^2 + M \mu_M^2, 
\label{N-M-1:Menergy}
\end{equation}
where the rest energy is
\begin{equation}
\omega^0_{N-M-1:M} = \frac{\bar{c}^2}{4} (N-1) M (N-M-1).
\end{equation}
The total momentum is the sum of the two string momenta,
\begin{equation}
k = k_s + k_M = (N-M-1) \mu_s + M \mu_M,
\label{N-M-1:Mmomentum}
\end{equation}
so we can write the energy as 
\begin{equation}
\omega_{N-M-1:M} (k_s, k_M) = \omega^0_{N-M-1:M} + \frac{k_s^2}{N-M-1} + \frac{k_M^2}{M}.
\end{equation}
The Bethe equations are here
\begin{eqnarray}
(N-M-1) \mu_s L - \Phi_{N-M-1,M} (\mu_s - \mu_M) = 2\pi I_s, \nonumber \\
M \mu_M L + \Phi_{N-M-1,M} (\mu_s - \mu_M) = 2\pi I_M,
\end{eqnarray}
with $I_s, I_M$ integers.  In the limit of large L, we again ignore the scattering phase shift,
and take $\mu_s$ and $\mu_M$ as free parameters.  The 
energy is again bounded from below by $\omega^l_{N-M-1:M} (k) = \omega^0_{N-M-1:M} + {k^2}/{N}$.
The dynamical constraints give in this case
\begin{eqnarray}
\fl
\mu_s^{\pm}(k,\omega) &=& \frac{k}{N-1} \mp \left[\frac{M}{(N-1)(N-M-1)}\right]^{1/2} [\omega - \omega^l_{N-M-1:M}(k)]^{1/2}, \nonumber \\
\fl
\mu_M^{\pm}(k,\omega) &=& \frac{k}{N-1} \pm \left[\frac{N-M-1}{(N-1)M}\right]^{1/2} [\omega - \omega^l_{N-M-1:M}(k)]^{1/2}.
\label{N-M-1:MmusmuM}
\end{eqnarray}
Therefore, in the large $L$ limit, these states again form a two-fold degenerate continuum 
beginning at the lower threshold $\omega^l_{N-M-1:M} (k)$ and extending to arbitrarily high energy,
\begin{equation}
\mbox{N-M-1:M continuum:} \hspace{1cm} \omega^l_{N-M-1:M}(k) \leq \omega < \infty.
\label{N-M-1:Mcontinuum}
\end{equation}

\begin{figure}
\begin{tabular}{cc}
\includegraphics[width=6cm]{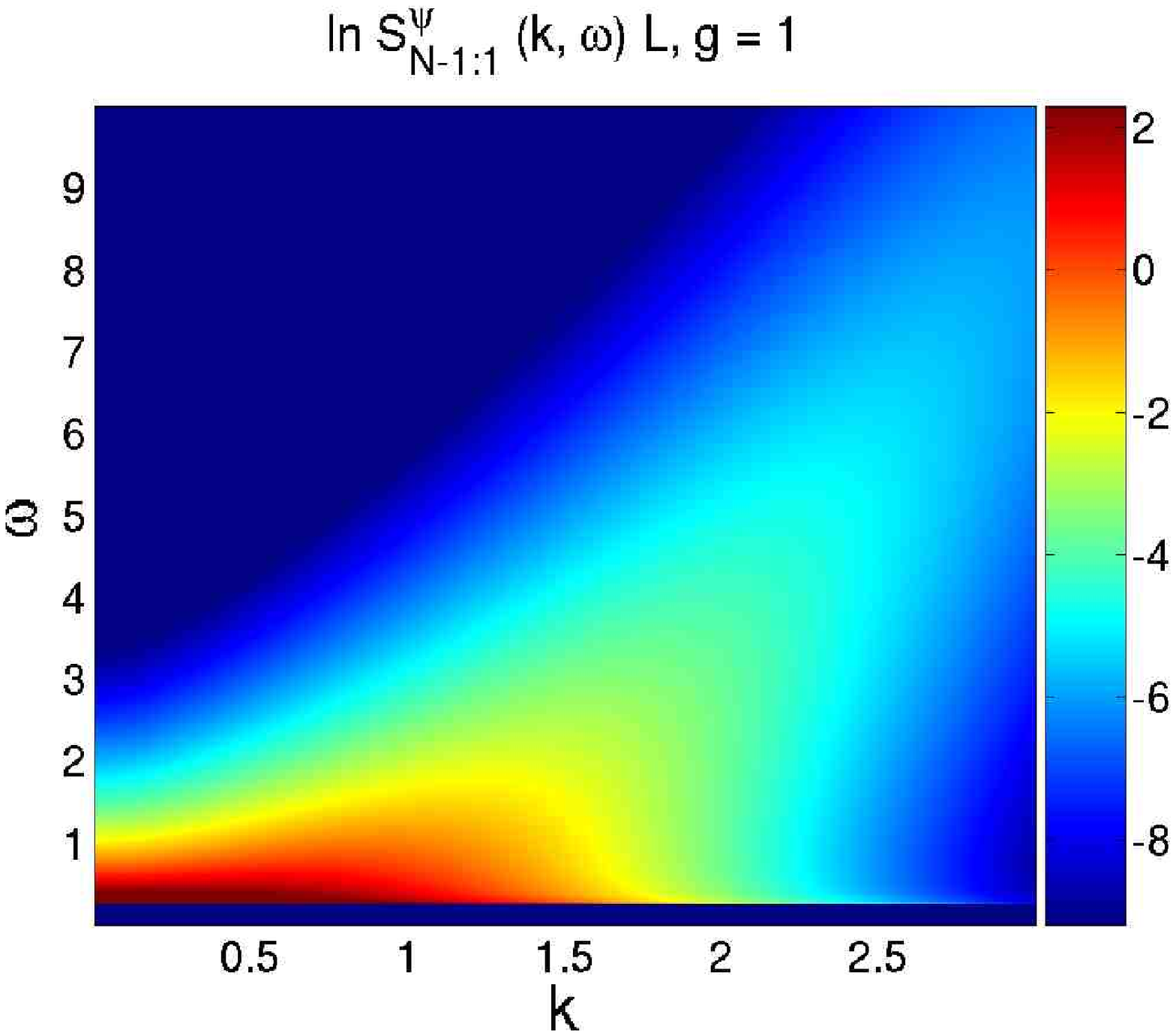} &
\includegraphics[width=6cm]{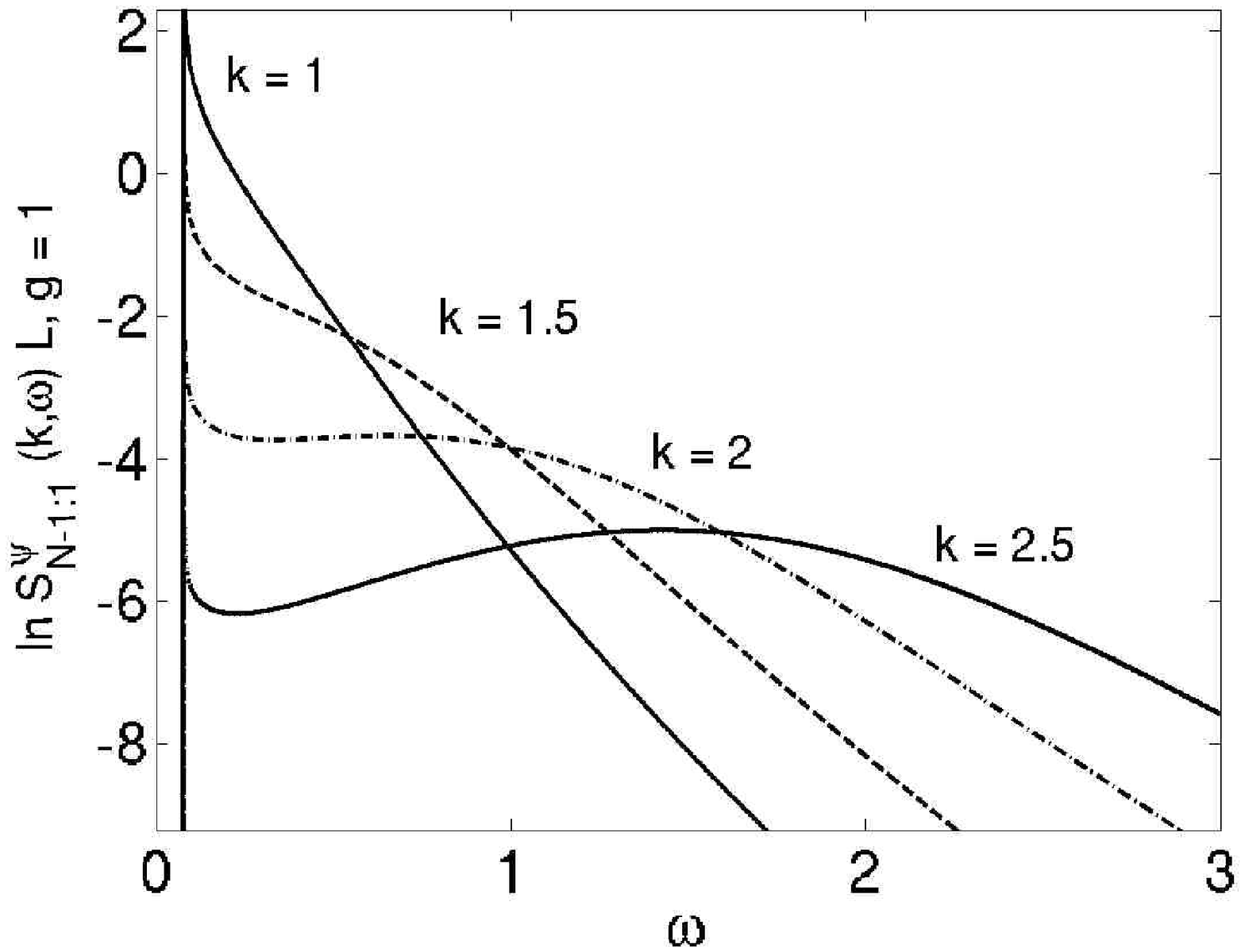} 
\end{tabular}
\caption{Left: contribution to the logarithm of the one-body function coming from $N-2:1$ (two-particle) states,
in the large $N$ limit for $g = 1$.  
Right:  Fixed momentum cuts of the same, showing the same qualitative features as for the DSF, 
namely the weakening of
the square-root singularity at higher momentum, the appearance of a local maximum as a function of frequency for
larger values of momentum, and the displacement of this
maximum towards higher energies at higher momentum.}
\label{Spsifig}
\end{figure}

The calculation of the contribution of theses states to the one-body function follows the same
logic as that used for the $N-M:M$ states for the structure factor.  We find
\begin{eqnarray}
\fl
S^{\Psi}_{N-M-1:M} (k,\omega) = \frac{\Theta(\omega -  \omega^l_{N-M-1:M}(k))}{\left[\omega - \omega^l_{N-M-1:M}(k)\right]^{1/2}}
\times \nonumber \\
\fl \hspace{1cm} \times \frac{\Gamma^4(N)}{2 \bar{c} L (N-1)^{1/2} [(N-M-1)M]^{3/2}} 
\sum_{\sigma = \pm}
\frac{H_{N-M-1}^2 (\mu_s^{\sigma}/\bar{c}) H_{M}^2 (\mu_M^{\sigma}/\bar{c})}{F_{N-M-1,M} (\mu_{s}^{\sigma} - \mu_M^{\sigma})}.
\end{eqnarray}
At large $N$, this can be simplified to
\begin{eqnarray}
\fl
S^{\Psi}_{N-M-1:M} (k,\omega) = \frac{\Theta(\omega -  \omega^l_{N-M-1:M}(k))}{\left[\omega - \omega^l_{N-M-1:M}(k)\right]^{1/2}}
\frac{N^{2-2M}}{2 g^2 L M^{3/2}} \sum_{\sigma} \frac{F^2_{M+1} (\mu^{\sigma}_s/\bar{c})}{\left[1/4 + (\mu^{\sigma_M}/g)^2\right]^{2M}}.
\end{eqnarray}
The leading $M=1$ term then becomes in this limit
\begin{eqnarray}
S^{\Psi}_{N-2:1} (k, \omega) = \frac{g^2}{2L} \frac{\Theta(\omega - g^2/4)}{[\omega - g^2/4]^{1/2} \omega^2}
\sum_{\sigma = \pm} F_2^2 (x_{\sigma})
\end{eqnarray}
with again $x_{\pm} = {\pi} (\sqrt{\omega - g^2/4} \pm k)/g$.  Here, $F_2 (x) = {\pi x}/{\sinh \pi x}$.
This contribution is plotted for $g = 1$ in Figure \ref{Spsifig}.

Note that the relative contribution of these states to the sumrule
(\ref{opsumrule}) goes like $O(N^0)/L$ and so it is suppressed by a factor $N$
compared to the one-particle state.  

\subsection{The momentum distribution function}

As for the structure factor, the static limit, known as momentum distribution 
function, 
\begin{equation}
S^{\Psi} (k) = 
\int_0^{\infty} \frac{d\omega}{2\pi} S^{\Psi} (k, \omega),
\label{static_GF}
\end{equation}
is obtained easily integrating the previous results.

The one-particle contribution is 
\begin{equation}
S^{\Psi}_{N-1} (k)=
\int_0^\infty \frac{d\omega}{2\pi} S^{\Psi}_{N-1} (k, \omega) = 
\frac{N}{gL} \frac{\pi^2}{\cosh^2 (\pi k/g)}\,,
\end{equation}  
that again decays exponentially for large $k$.

\begin{SCfigure}[50]
\includegraphics[width=6cm]{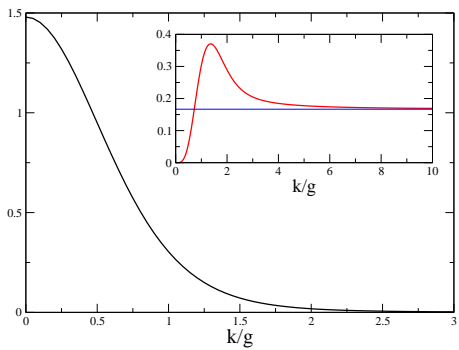} 
\caption{The two-particle contribution to the momentum distribution function
$g L S^{\Psi}_{N-2:1} (k)$.
Inset: $L k^4 S^{\Psi}_{N-2:1} (k)/g^3$ to show the $k^{-4}$ 
tail with prefactor $1/6$ (straight line).}
\label{staticop}
\end{SCfigure}

We can also evaluate the contribution of the $N-2:1$ states to the static
one-body function for which we get the integral
\begin{equation}
S^{\Psi}_{N-2:1} (k) = \frac{\pi^2}{gL} \int_{-\infty}^{\infty}
\frac{d\zeta}{2\pi} \frac{1}{(\zeta^2 + 1/4)^2} \frac{(\zeta - k/g)^2}{\sinh^2 \pi(\zeta - k/g)}.
\end{equation}
This integral can be done summing over all the residues. The final result
involves polygamma functions and it is not really illuminating. For this 
reason we only plot it in Fig. \ref{staticop}.

Let instead study in detail the behavior for large $k$. In this case we have
\begin{equation}
S^{\Psi}_{N-2:1} (k\gg g) = \frac{\pi^2}{L} \frac{g^3}{k^4}\int_{-\infty}^{\infty}
\frac{d\zeta}{2\pi} \frac{\zeta^2}{\sinh^2 \pi\zeta}=
\frac{g^3}{6L k^4}.
\label{k4attr}
\end{equation}
This $k^{-4}$ behaviour is the same as for the repulsive regime \cite{OlshaniiPRL91,MinguzziPLA294}. 
In fact this is not a coincidence: 
The derivation of Ref. \cite{OlshaniiPRL91,MinguzziPLA294} relating the large $k$ behavior of 
$S^{\Psi} (k)$ to the second density moment $\langle \rho^2 \rangle$ makes use
only of analytical properties of the Bethe wave function that hold both 
in attractive and repulsive regimes. 
Adapting the result of Ref.  \cite{OlshaniiPRL91,MinguzziPLA294} to our normalization we get
\begin{equation}
S^{\Psi} (k\to\infty)= \langle \rho^2 \rangle \frac{c^2}{k^4}\,,
\end{equation}
and inserting the value of $\langle \rho^2 \rangle$ obtained via the  
Hellmann-Feynman theorem in Eq. (\ref{HF}) reproduces exactly 
our last result (\ref{k4attr}).
This is not a trivial result because it is saying that the large 
$k$ behaviour is completely determined by the two-particle states.

\section{Conclusions and perspective}
\label{sec:conclusion}
We presented a detailed calculation of the zero-temperature dynamical
correlation functions of the attractive Lieb-Liniger model combining the
determinant representation of the matrix elements with the exact (with
exponential precision in the system size) knowledge of the Bethe states.  
All the sum rules we have studied are essentially saturated by the very 
simplest classes of excitations: 
the one-particle states obtained by giving finite momentum to the ground
state string (this mode corresponds to the Goldstone mode associated to 
spontaneous symmetry breaking in the Gross-Pitaevskii framework), or the 
two-particle states obtained when splitting it up by extracting a single atom.
The response of the system to an external probe therefore essentially occurs
with zero recoil energy, which is interpreted as a M\"ossbauer-like effect:
the attractive gas is really like a crystal along the imaginary axis in
momentum space, and responds like a single particle since the gap to the lowest
excitations is finite.  
For large atom number $N$, higher families of excited states have their 
contribution to the correlation functions suppressed by progressively higher 
powers of $1/N$, and can be ignored for all practical purposes.  
Our results thus provide an extremely accurate, essentially exact
picture of the zero-temperature dynamics of the system.

Using our results, it is possible to compute other more elaborate form factors, for
example those of the density or field operator between higher excited states involving
nontrivial string partitioning.  These take the form of reduced determinants involving only
the string centers, but are still obtainable using a reasoning similar to the one used in the
calculation of the single-string form factors in \ref{app:density_FF} and \ref{app:field_FF}.
We have not pursued this line here since these would contribute sub-dominantly to the 
observable zero-temperature lineshapes.  

Integrable models for atomic gases also include the class of multi-species, $\delta$-interaction
particles.  For these, the Bethe Ansatz involves a nesting, making the current approach somewhat
more complicated.  It would be interesting to see whether the methods used here could be adapted to
such cases, and used to derive related expressions for correlation functions like the spin 
structure factor.  Such an object would also be in principle observable experimentally
\cite{CarusottoJPB39}.

Another interesting possibility is to consider the effects of a finite temperature.
This would probably be of great importance for experiments, since the ground state is part
of a dispersionless band in the large $N$ limit;  any temperature greater than $\mbox{O}(1/N)$
would thus quickly smear out the correlations in the momentum direction.  However, all the necessary
form factors for the calculation of the finite-temperature response functions are readily
obtained directly from the results we presented, and it is more or less straightforward to
obtain the finite temperature dynamical correlations.  
The presence of a harmonic trap can also be taken into account by convolving the ground state
wavepacket with a harmonic oscillator wavefunction.  
We will investigate these issues further
in the near future.

In conclusion, we have shown that the attractive Lieb-Liniger gas is a system for
which the Bethe Ansatz is sufficiently tractable to allow an analytical calculation
of important correlation functions.

\ack
Both authors acknowledge stimulating discussions with G. Mussardo and 
F. H. L. Essler in the early phase of this work, and are thankful for support 
from the Stichting voor Fundamenteel Onderzoek der Materie (FOM) in 
the Netherlands.
P. C. further gratefully acknowledges the ESF exchange grant 
1311 of the INSTANS European Network.
J.-S. C. acknowledges interesting discussions with J. Brand, I. Carusotto, Y. Castin, A. Yu. Cherny and G. V. Shlyapnikov
during a stay at the Institut Henri Poincar\'e-Centre \'Emile Borel, which is thanked for hospitality and support.

\appendix

\section{Reduced Bethe equations}
\label{app:Reduced_BE}
In this appendix we derive the reduced Bethe equations for the string centers. 
The Bethe equation (\ref{BE}) in terms of the string rapidities (\ref{string_rap}) 
are
\begin{eqnarray}
\fl e^{i \lambda^{j,a}_{\alpha} L} = \prod_{(k, \beta, b) \neq (j, \alpha, a)} 
\frac{\lambda^{j, a}_{\alpha} - \lambda^{k, b}_{\beta} - i\bar{c}}{\lambda^{j, a}_{\alpha} - \lambda^{k, b}_{\beta} + i\bar{c}} 
\nonumber \\
\fl \hspace{1cm} = \prod_{(k, \beta) \neq (j, \alpha)} \prod_{b=1}^k 
\frac{\lambda^j_{\alpha} - \lambda^k_{\beta} + i\bar{c} (\frac{j-k}{2} - a + b - 1)}
{\lambda^j_{\alpha} - \lambda^k_{\beta} + i\bar{c} (\frac{j-k}{2} - a + b + 1)}
\prod_{b \neq a} \frac{\bar{c} (-a + b - 1) + \delta^{j, (a,b)}_{\alpha}}{\bar{c} (-a + b + 1) + \delta^{j, (a,b)}_{\alpha}}
\label{Reduced_BE_step1}
\end{eqnarray}
where we have separated inter- and intra-string parts, dropped all string deviations for inter-string factors, 
and denoted $\delta^{j, a}_{\alpha} - \delta^{j, b}_{\alpha} = \delta^{j, (a,b)}_{\alpha}$ in the intra-string part.
Simplified Bethe equations are obtained by taking the product of these equations within the string
considered.  The left-hand side becomes
\begin{equation}
\prod_{a=1}^j e^{i \lambda^{j,a}_{\alpha} L} = e^{i j\lambda^j_{\alpha} L}.
\end{equation}
On the right-hand side, we have
\begin{equation}
\prod_{a=1}^j \prod_{b \neq a} \frac{\bar{c} (-a + b - 1) + \delta^{j, (a,b)}_{\alpha}}{\bar{c} (-a + b + 1) + \delta^{j, (a,b)}_{\alpha}}
= (-1)^{j(j+1)} = 1
\end{equation}
for the intra-string part, and (writing $\lambda = \lambda^j_{\alpha} - \lambda^k_{\beta}$)
\begin{eqnarray}
\prod_{a=1}^j \prod_{b=1}^k 
\frac{\lambda + i\bar{c} (\frac{j-k}{2} - a + b - 1)}
{\lambda + i\bar{c} (\frac{j-k}{2} - a + b + 1)} \nonumber \\
= e_{|j - k|} (\lambda) e_{|j-k| + 2}^2 (\lambda) e_{|j - k| + 4}^2 (\lambda) ... e_{j+k - 2}^2 (\lambda) e_{j+k} (\lambda)
\equiv E_{jk} (\lambda)
\end{eqnarray}
where 
\begin{equation}
e_j (\lambda) = \frac{\lambda - i\bar{c}j/2}{\lambda + i\bar{c}j/2}.
\end{equation}
The exponential form of the Bethe equations has thus been reduced to the set of 
$N_s$ coupled equations for the string centers $\lambda^j_{\alpha}$,
\begin{equation}
e^{i j \lambda^j_{\alpha}L} = \prod_{(k, \beta) \neq (j, \alpha)} E_{jk} (\lambda^j_{\alpha} - \lambda^k_{\beta}).
\end{equation}
Taking the logarithm and defining
\begin{equation}
\phi_j (\lambda) = 2 ~\mbox{atan}~ \frac{2\lambda}{\bar{c} j}
\end{equation}
such that $-i \log (-e_j (\lambda)) = \phi_j(\lambda)$, we find equations
(\ref{Reduced_BE}) and (\ref{Reduced_Phi}).

\section{Norm of a general state}
\label{app:norm}
In this appendix, we consider the fate of the Gaudin-Korepin norm formula in the case
of general string states.  A similar treatment can be found in \cite{KirillovJMS40}.
First of all, the prefactor of equation (\ref{usual_norm}) explicitly reads (in string notation)
\begin{eqnarray}
\fl
\prod_{(k,\beta,b)\neq(j,\alpha,a)} \!\!\!\!\frac{\lambda^{j,a}_{\alpha} - \lambda^{k,b}_{\beta} + i\bar{c}}{\lambda^{j,a}_{\alpha} - \lambda^{k,b}_{\beta}}
= \!\!\!\! \prod_{(k, \beta)\neq (j,\alpha)} \prod_{a=1}^j \prod_{b=1}^k \frac{\lambda^j_{\alpha} - \lambda^k_{\beta} + i\bar{c} (\frac{j-k}{2} -a+b+1)}
{\lambda^j_{\alpha} - \lambda^k_{\beta} + i\bar{c} (\frac{j-k}{2} -a+b)}
\nonumber \\
\hspace{4cm}\times \prod_{j,\alpha} \prod_{a=1}^j \prod_{b\neq a} \frac{\bar{c} (-a+b+1) + \delta^{j,(a,b)}_{\alpha}}{\bar{c} (-a +b)}
\label{norm_prefactor_1}
\end{eqnarray}
where we have separated the product into inter- and intra-string parts and 
defined $\delta^{j,(a,b)}_{\alpha} = \delta^{j,a}_{\alpha} - \delta^{j,b}_{\alpha}$.

To leading order in string deviations, this prefactor is then simplified to
(including the $\bar{c}^N$ factor)
\begin{eqnarray}
\fl
\bar{c}^{N_s} \prod_{j} j^{N_j} 
\prod_{(k, \beta) > (j, \alpha)} F_{jk}(\lambda_{\alpha}^j - \lambda_{\beta}^k) \times \prod_{j,\alpha} \prod_{a=1}^{j-1} \delta^{j,(a,a+1)}_{\alpha}
\times (1 + \mbox{O}(\delta))
\label{norm_prefactor}
\end{eqnarray}
where 
\begin{eqnarray}
F_{jk} (\lambda) = \frac{\lambda^2 + (\frac{\bar{c}}{2} (j + k))^2}
{\lambda^2 + (\frac{\bar{c}}{2} (j - k))^2}.
\end{eqnarray}

Let us now deal with the Gaudin matrix.  In string notation, the matrix elements read
\begin{eqnarray}
\fl
{\cal G}_{(j,\alpha,a) (k,\beta,b)} = \delta_{(j,\alpha,a) (k,\beta,b)} \left[L + \sum_{(l,\gamma,c)} K_{(j,\alpha,a) (l,\gamma,c)}\right] 
- K_{(j,\alpha,a) (k,\beta,b)}
\end{eqnarray}
where 
\begin{eqnarray}
K_{(j,\alpha,a) (k,\beta,b)} = -\frac{2\bar{c}}{(\lambda^{j,a}_{\alpha} - \lambda^{k,b}_{\beta})^2 + \bar{c}^2}.
\end{eqnarray}
The diagonal block associated to a particular string within the Gaudin matrix is
\begin{eqnarray}
\fl
\scriptsize
\left(
\begin{array}{ccccc}
L + K_{12} + S_1 & -K_{12} & -K_{13} & ... & -K_{1j} \\
-K_{12} & L + K_{12} + K_{23} + S_2 & -K_{23} & ... & -K_{2j} \\
-K_{13} & -K_{23} & L + K_{23} + K_{34} + S_3 & ... & ... \\
... & ... & ... & ... & -K_{j-1 j} \\
-K_{1j} & ... & ... & -K_{j-1 j} & L + K_{j-1 j} + S_j 
\end{array} \right)
\end{eqnarray}
\normalsize
in which we have isolated elements of the type $K_{a a+1}$ which are exponentially large,
\begin{eqnarray}
K_{a a+1} = -\frac{2\bar{c}}{(i\delta^{(a,a+1)} + i\bar{c})^2 + \bar{c}^2} = \frac{1}{\delta^{(a,a+1)}} (1 + \mbox{O}(\delta)).
\end{eqnarray}
All other elements, including the sums $S_j$ (of finite $K_{ab}$ terms) are of zeroth order in
string deviations.  

The determinant can now be manipulated by adding block column 1 to block column 2, followed by
adding block row 1 to block row 2.  This eliminates the $K_{12}$ factor except in the block $(1,1)$
position.  Factors $K_{a a+1}$ for $a = 2, ..., j-1$ can be similarly eliminated from two off-diagonal
and one diagonal positions.  The string block then has $j-1$ exponentially large 
factors $K_{a a+1}$ on the diagonal, and all off-diagonal terms within the block can be neglected.  
Doing similar operations for all strings in the eigenstate and keeping all remaining factors
into account, we find that, 
to leading order in the string deviations, the determinant of the Gaudin matrix becomes 
\begin{eqnarray}
\mbox{Det} {\cal G} = \prod_{j,\alpha} \prod_{a=1}^{j-1} \frac{1}{\delta^{j,(a,a+1)}_{\alpha}}
\times \mbox{Det} {\cal G}^{(r)} \times (1 + \mbox{O}(\delta))
\end{eqnarray}
where the reduced Gaudin matrix is an $N_s \times N_s$ matrix whose elements are
\begin{eqnarray}
{\cal G}^{(r)}_{(j, \alpha), (k, \beta)} = \frac{\partial}{\partial \lambda_{\beta}^k} \left[ j \lambda_{\alpha}^j L 
- \sum_{(l, \gamma)} \Phi_{jl} (\lambda_{\alpha}^j - \lambda_{\gamma}^l) \right].
\end{eqnarray}

Upon taking the product of this with the prefactor (\ref{norm_prefactor}), the
string deviation products cancel, and we find that the  
norm of a completely general string state (dropping corrections of order $\delta$)
is finite and can be written as
\begin{eqnarray}
||\Psi_N (\{ \lambda_{\alpha}^j \}) ||^2 = \bar{c}^{N_s} \prod_{j} j^{N_j} 
\prod_{(k, \beta) > (j, \alpha)} F_{jk}(\lambda_{\alpha}^j - \lambda_{\beta}^k)
\hspace{0.2cm}\mbox{Det} {\cal G}^{(r)}.
\end{eqnarray}
In the limit of large $L$, we can simplify the reduced Gaudin matrix to 
${\cal G}^{(r)}_{(j,\alpha), (k,\beta)} = j L \delta_{jk} \delta_{\alpha \beta} $,
allowing to simplify the norm to (\ref{string_state_norm}).

\section{Density operator form factor}
\label{app:density_FF}
The form factor of the density operator between two Bethe eigenstates of the Bose gas is
given in \cite{SlavnovTMP79,SlavnovTMP82} as 
\be
\fl
\Sigma^{\rho}_{\mu \lambda}= (P_{\mu} - P_{\lambda}) \prod_j(V_j^+ -V^-_j)\prod_{j,k} 
\left(\frac{\l_{jk}+ic}{\mu_j-\l_k}\right)
\mbox{Det}_N [{\bf I}+\bar{\bf U}(\l_p)]\frac1{V_p^+ -V^-_p}\,,
\ee
with
\be
V^\pm_j=\prod_m\frac{\mu_m-\l_j\pm ic}{\l_m-\l_j\pm ic}\,,
\ee
and
\be
\bar{U}_{jk}(\l_p)=i\frac{\mu_j-\l_j}{V_j^+ -V^-_j}\prod_{m\neq j}
\left(\frac{\mu_m-\l_j}{\l_m-\l_j}\right)
(K (\lambda_j - \lambda_k)-K (\lambda_p- \lambda_k))\,,
\ee
and where the kernel $K$ is given by Eq. (\ref{kernel})
and $\l_p$ an arbitrary number, not necessarily a rapidity.

Fixing $\l_p=\l_N$, the last column ($k=N$) becomes $(0,\dots,0,1)$ and the form 
factor can be written in terms of an $(N-1)$ determinant:
\begin{equation}
\Sigma^{\rho}_{\mu \lambda} = i^{N-1}  (P_{\mu} - P_{\lambda})\prod_{j,k} \frac{\lambda_{jk} + ic}{\mu_j - \lambda_k} 
~\mbox{Det}_{N-1} U (\{\mu\}, \{ \lambda \})
\end{equation}
with the matrix $U$ having elements
\begin{eqnarray}
\fl
U_{jk} (\{ \mu \}, \{ \lambda \}) = \delta_{jk} \frac{V^+_j - V^-_j}{i} + \frac{\prod_{a=1}^{N} (\mu_a - \lambda_j)}
{\prod_{a\neq j}^{N} (\lambda_a - \lambda_j)} (K(\lambda_j - \lambda_k) - K (\lambda_{N} - \lambda_k)).
\label{U}
\end{eqnarray}

Let us now consider the case where rapidities $\{ \lambda\}$ form an $N$-string, as for
the ground state.  We leave the set of rapidities $\{ \mu \}$ arbitrary.  We thus take
a simple notation in which
\begin{equation}
\lambda_j = \lambda + \frac{i\bar{c}}{2} (N + 1 - 2j) + i \delta_j.
\label{GS_string}
\end{equation}
We have to deal explicitly with the small deviations $\delta_j$ in our matrices.
Namely, we have 
\begin{eqnarray}
K (\lambda_j - \lambda_k) = \left\{ \begin{array}{cc}
\displaystyle
\frac{1}{\delta_{j j+1}} + \mbox{O}(1), & k = j+1, \\ &\\ \displaystyle
\frac{1}{\delta_{j-1 j}} + \mbox{O}(1), & k = j-1, \\ & \\
\mbox{O}(1), & k \neq j \pm 1.
\end{array} \right.
\end{eqnarray}
In these, we have written $\delta_{jk} = \delta_j - \delta_k$ and kept only the leading term,
which turns out to be the only one needed for the calculation.  

We also define the regular function
\begin{equation}
V_j = \frac{\prod_{m=1}^N (\mu_m - \lambda_j)}{\prod_{m\neq j}^N (\lambda_m - \lambda_j)}, 
\hspace{1cm} j = 0, 1, ..., N+1
\end{equation}
by defining $\lambda_0$ and $\lambda_{N+1}$ using equation (\ref{GS_string}) for $j = 0, N+1$
(we don't need to define $\delta_j$ for these values of $j$).  The $V_j^{\pm}$ functions
can thus be written in regularized form
\begin{eqnarray}
V^+_j = \left\{ \begin{array}{cc}
V_0, & j = 1, \\ &\\ \displaystyle
\frac{-i}{\delta_{j-1 j}} V_{j-1}, & j = 2, ..., N,
\end{array} \right. \nonumber \\
V^-_j = \left\{ \begin{array}{cc}\displaystyle
\frac{i}{\delta_{j j+1}} V_{j+1}, & j = 1, ..., N - 1, \\&\\
V_{N+1}, & j = N.
\end{array} \right.
\end{eqnarray}

To leading order, the $U$ matrix (\ref{U}) is 
\be
\fl \tiny
\left(
\begin{array}{ccccccc}
\frac{-V_2}{\d_{12}}& \frac{V_1}{\d_{12}}&0&0&\dots& 0&\frac{-V_1}{\d_{N-1 N}}\\
\frac{V_2}{\d_{12}}&\frac{-V_1}{\d_{12}} -\frac{V_3}{\d_{23}}&\frac{V_2}{\d_{23}}&0
&\dots& 0&\frac{-V_2}{\d_{N-1 N}}\\
0&\frac{V_3}{\d_{23}}&\frac{-V_2}{\d_{23}}-\frac{V_4}{\d_{34}}&
\frac{V_3}{\d_{34}}&\dots&0& \frac{-V_3}{\d_{N-1}}\\
\vdots&\vdots&\vdots&\vdots&\ddots&\vdots&\vdots\\
0&0&0&0&\dots&\frac{-V_{N-3}}{\d_{N-4 N-3}}-\frac{V_{N-1}}{\d_{N-2 N-1}}&
\frac{V_{N-2}}{\d_{N-2 N-1}} - \frac{V_{N-2}}{\d_{N-1 N}}\\
0&0&0&0&\dots&\frac{V_{N-1}}{\d_{N-2 N-1}}&
\frac{-V_{N-2}}{\d_{N-2 N-1}}-\frac{V_{N-1} + V_N}{\d_{N-1 N}}
\end{array}\right)
\ee
The determinant can be evaluated by means of simple row and column manipulations.
Adding the first row to the second one, thereafter the second to the third, as so on,
we end up with an upper triangular matrix whose determinant is simply
\begin{equation}
\mbox{Det} ~U = (-1)^{N-1}\left[\prod_{j=1}^{N-1} \frac{1}{\delta_{j j+1}}\right] \left[\prod_{j=2}^{N-2} V_j\right] 
\left[\sum_{j=1}^N V_j\right].
\end{equation}
Using $\prod_{m \neq j} (\lambda_m - \lambda_j) = (i\bar{c})^{N-1} \prod_{m\neq j} (-m + j) = (i\bar{c})^{N-1} (-1)^{N-j} (j-1)! (N-j)!$
for the denominator of $V_j$ and defining the function
\begin{eqnarray}
Q_j = \prod_m (\mu_m - \lambda_j)
\end{eqnarray}
for the numerator, we arrive after simple manipulations to the following expression for
the density operator form factor (taking the ground state momentum to zero)
\begin{equation}
\fl
\Sigma^{\rho}_{\mu \lambda} = \!\frac{[i\bar{c}]^N P_{\mu} N!}{Q_1 Q_N}
\sum_{j=1}^N (-1)^{j-1} \!\! \left( \!\! \begin{array}{c} N-1 \\ j-1 \end{array} \!\!\right) Q_j 
= \frac{\bar{c}^{2N} P_{\mu}N!}{2^N Q_1 Q_N} S_{N,N} (\{ 2i \mu_m /\bar{c}\})
\end{equation}
and where we have defined the summation
\begin{equation}
S_{N,M} (\{ \mu \}_M) = \sum_{j=1}^N (-1)^{j-1} \left( \begin{array}{c} N-1 \\ j-1 \end{array} \right)
\prod_{m=1}^M \left[ N + 1 - 2j + \mu_{m} \right]
\end{equation}
for any set of complex numbers $\{ \mu \}$.  This sum can be performed explicitly
(see \ref{app:summation}), yielding the final expression for the density operator form factor
between the ground state ($N$-string centered on zero) and the arbitrary string state
defined by the set of string rapidities $\{ \mu^j_{\alpha} \}$:
\begin{equation}
\Sigma^{\rho}_{\mu \lambda} = \frac{N \Gamma^2 (N) P_{\mu}^2/\bar{c}}{\prod_{j, \alpha, a} \left[ (\mu_{\alpha}^j/\bar{c})^2 
+ (\frac{N+j}{2} - a)^2 \right]}.
\label{density_FF_app_res}
\end{equation}

\section{Field operator form factor}
\label{app:field_FF}
The form factor of the field operator between two Bethe states is \cite{KojimaCMP188,CauxJSTATP01008}
\be
\Sigma^{\Psi}_{\mu \lambda} = i^{N-1} c^{-1/2} \frac{\prod_{j,k}^N \l_{jk}+ ic}{
\prod_{j}^N \prod_k^{N-1} \mu_k - \l_j}
\mbox{Det}_{N-1} U (\{\mu \}, \{ \lambda \})\,,
\ee
with the matrix $U$ once again given by (\ref{U}), but with the difference
that the set of rapidities $\mu$ now comprises $N-1$ elements.
The calculation is almost identical to that for the density operator, 
except that here we have 
$Q_m=\prod_{j=1}^{N-1}(\mu_j-\l_m)$, i.e. it is the product of $N-1$ and not 
$N$ factors.  The prefactor is slightly different, and we get
\begin{equation*}
\fl
\Sigma^{\Psi}_{\mu \lambda} = \frac{i^{N-1}\bar{c}^{N - 1/2} N!}{Q_1 Q_N} \sum_{j=1}^N (-1)^{j-1}
\left( \!\! \begin{array}{c} N-1 \\ j-1 \end{array} \!\!\right) Q_j
= \frac{\bar{c}^{2N - 3/2} N!}{2^{N-1} Q_1 Q_N} S_{N,N-1} (\{2i \mu_j/\bar{c} \}).
\end{equation*}
The summation can also be performed explicitly for this new definition of $Q_j$
using the general results of \ref{app:summation}, we finally find
\be
\Sigma^{\Psi}_{\mu \lambda} = \frac{\bar{c}^{1/2} N \Gamma^2 (N)}
{\prod_{j, \alpha, a} \left[ (\mu_{\alpha}^j/\bar{c})^2 
+ (\frac{N+j}{2} - a)^2 \right]}.
\label{field_FF_app_res}
\ee
Note once again that the set of $\{ \mu \}$ rapidities comprises $N-1$ elements 
instead of $N$ like in (\ref{density_FF_app_res}), so although these terms look
the same, the different $\mu$ string contents involved make them different.

\section{Proof of summation formula}
\label{app:summation}
We wish to compute
\begin{equation}
S_{N,M} (\{ \mu \}_M) = \sum_{j=1}^N (-1)^{j-1} \left( \begin{array}{c} N-1 \\ j-1 \end{array} \right)
\prod_{m=1}^M \left[ N + 1 - 2j + \mu_{m} \right]
\end{equation}
for $M \leq N$ and an arbitrary set of parameters $\{ \mu \}$.  The product can be written as
\begin{equation}
\prod_{m=1}^M \left[ N + 1 - 2j + \mu_{m} \right] = \sum_{l = 0}^M (N+1-2j)^{M-l} P_l (\{ \mu \})
\end{equation}
where $P_l$ is the $l$-th order completely symmetric polynomial.  We therefore need to compute all
\begin{equation}
T_{N,L} = \sum_{j=1}^N (-1)^{j-1} \left( \begin{array}{c} N-1 \\ j-1 \end{array} \right)
(N+1-2j)^L, \hspace{0.5cm} L = 0, ..., N.
\end{equation}
Consider 
\begin{equation}
\fl
U_N (a) = \sum_{j=0}^{N-1} (-1)^j \left( \begin{array}{c} N-1 \\ j \end{array} \right) e^{a(N-1-2j)}
= e^{a(N-1)} (1 - e^{-2a})^{N-1} = (2 \sinh a)^{N-1}.
\end{equation}
This is a generating function for $T_{N,L}$ since
\begin{equation}
T_{N,L} = \frac{\partial^L}{\partial a^L} U_N (a) |_{a=0}.
\end{equation}
For the summations we need in order to calculate the density and field operator form factors, 
we can keep only the leading order in $a$ and write $U_N(a) = 2^{N-1} a^{N-1} + ...$.
Therefore, we have
\begin{equation}
T_{N,L} = \delta_{L, N-1} 2^{N-1} (N-1)! 
\end{equation}
which yields
\begin{equation}
S_{N,M} = 2^{N-1} (N-1)! \left(\delta_{M,N} P_1 (\{ \mu \}) + \delta_{M, N-1} \right).
\end{equation}
This can now be directly used in the form factor calculations of \ref{app:density_FF}
and \ref{app:field_FF}.


\end{document}